\documentclass[reprint,
groupedaddress,
showpacs,preprintnumbers,
 amsmath,amssymb,
 aps,
prb
]{revtex4-1}
\usepackage{graphicx}
\usepackage{dcolumn}
\usepackage{bm}
\usepackage{mathrsfs}
\begin{document}


\title{The Josephson current in Fe-based superconducting junctions: theory and experiment.}
\author{A.\,V.~Burmistrova}
\affiliation{Lomonosov Moscow State University, Faculty of Physics, 1(2), Leninskie gory, GSP-1, Moscow 119991, Russian Federation}
\affiliation{Lomonosov Moscow State University Skobeltsyn Institute of Nuclear Physics, 1(2), Leninskie gory, GSP-1, Moscow 119991, Russian Federation}
\affiliation{Moscow Institute of Physics and Technology, Dolgoprudny, Moscow 141700, Russia}
\affiliation{Moscow State Pedagogical University, Moscow 119992, Russian Federation}
\author{ I.\,A.~Devyatov}
\email[]{igor-devyatov@yandex.ru}
\affiliation{Lomonosov Moscow State University Skobeltsyn Institute of Nuclear Physics, 1(2),  Leninskie gory, GSP-1, Moscow 119991, Russian Federation}
\affiliation{Moscow Institute of Physics and Technology, Dolgoprudny, Moscow 141700, Russia}

\author{Alexander A. Golubov}
\affiliation{Faculty of Science and Technology and MESA+ Institute of Nanotechnology,
University of Twente, 7500 AE, Enschede, The Netherlands}
\affiliation{Moscow Institute of Physics and Technology, Dolgoprudny, Moscow 141700, Russia}

\author{Keiji Yada}
\affiliation{Department of Applied Physics, Nagoya University, Nagoya 464-8603, Japan}

\author{Yukio Tanaka}
\affiliation{Department of Applied Physics, Nagoya University, Nagoya 464-8603, Japan}
\affiliation{Moscow Institute of Physics and Technology, Dolgoprudny, Moscow 141700, Russia}

\author{M. Tortello and R.S. Gonnelli}
\affiliation{Dipartimento di Scienza Applicata e Tecnologia, Politecnico di Torino, 10129 Italy}

\author{V.A. Stepanov}
\affiliation{P. N. Lebedev Physical Institute, Russian Academy of Sciences, 119991, Moscow, Russia}

\author{Xiaxin Ding and Hai-Hu Wen}
\affiliation{Center for Superconducting Physics and Materials, National Laboratory of Solid State Microstructures and Department of Physics, Collaborative Innovation Center for Advanced Microstructures, Nanjing University, Nanjing 210093, China}

\author{L.H. Greene}
\affiliation{Department of Physics and the Material Research Laboratory, University of Illinois at
Urbana-Champaign, Urbana, Illinois 61801, USA}

\date{\today}

\begin{abstract}
We present theory of dc Josephson effect in contacts between Fe-based and spin-singlet $s$-wave superconductors. The method is based on the calculation of temperature Green's function in the junction within the tight-binding model. We calculate the phase dependencies of the Josephson current for different orientations of the junction  relative to the crystallographic axes of Fe-based superconductor. Further, we consider the dependence of the Josephson current on the thickness of an insulating layer and on temperature.  Experimental data for PbIn/Ba$_{1-x}$K$_{x}$(FeAs)$_2$ point-contact Josephson junctions are consistent with theoretical predictions for $s_{\pm}$ symmetry of an order parameter in this material. The proposed method can be further applied to calculations of the dc Josephson current in contacts with other new unconventional multiorbital superconductors, such as $Sr_2RuO_4$ and superconducting topological insulator $Cu_xBi_2Se_3$.
\end{abstract}

\pacs{74.20.Rp,74.70.Xa,74.45.+c,74.50.+r,74.55.+v}

\maketitle

\section{\label{sec1}Introduction}
An order parameter symmetry in unconventional superconductors contains an important information about superconducting pairing mechanism.
It is well-known that the phase-sensitive tunneling experiments
in junctions with unconventional superconductors
provide an important information about
 the symmetry of the order parameter
\cite{Wollman1993,Tsuei1994,VanHarlingen1995,Tsuei2000}.
Theory of quasiparticle tunneling spectroscopy
of a junction between  normal metal and  unconventional superconductor
was developed in \cite{tanaka1,kashiwaya00} and the existence of
midgap Andreev bound states was predicted \cite{Hu}.
Theory of Josephson current composed of unconventional superconductor junctions
was developed in \cite{Barash,tanaka97}.
After the discovery of high $T_{C}$ cuprates, several
new types of unconventional superconductors have been discovered.
The common property of these new unconventional superconductors
like Sr$_2$RuO$_4$\cite{Maeno1994,Mao2001,Kashiwaya11},
Fe-based superconductors (FeBS) \cite{kam}, and doped superconducting
insulators  Cu$_x$Bi$_2$Se$_3$ \cite{hor10,sasaki11} is
that all of them are  multiorbital materials.

All these materials have complex single-particle excitation spectrum, and one can expect sign-changing of superconducting order parameters in momentum space. The interband and intervalley scattering in these multiband unconventional superconductors significantly influences their energy spectrum.
Several phenomenological theories of transport in junctions based on FeBS
have been proposed in the past\cite{sacr,dev1,lind,gol,rom,kar1,kar2}.
However, only recently a microscopic theory of the quasiparticle current in
normal metal - multiband superconductor junctions was formulated \cite{bcrus,Burmistrova2013},
which takes into account unusual properties of these materials.
But there is still no microscopic theory to describe the Josephson
current in junctions between multiband superconductor and conventional
single-band spin-singlet $s$-wave superconductor, which goes
beyond the existing phenomenological theories
\cite{ber,chen,koshelev2012,lind}.

The aim of this paper is
to propose a microscopic theory of the Josephson current
in junctions based on multiband superconductors. We apply this theory to calculate  the
current phase relations in junctions
between spin-singlet $s$-wave and FeBS's for different
orientations. We also calculate the temperature dependencies of the critical Josephson current in these junctions. We confirm that the recently proposed
phase sensitive experiment \cite{Golubov2013}
is feasible to determine the symmetry of the order parameter in FeBS's.
Brief account of basic results of this paper is given in Ref. \cite{EPL2014}.

The organization of our paper is as follows.
In Sec. II we discuss the general formulation of our tight-binding approach  for the calculation of the dc Josephson current in junctions with single-orbital superconductors.
We demonstrate that our tight-binding approach
reproduces the previous results for the
Josephson effect in junctions with single-orbital superconductors.
In Sec. III we present the application of our method for the
calculation for the multiorbital case.
We consider FeBS in the framework of the two-band model and describe the
detailed procedure of the calculation of the Josephson current
for in-plane and out-of-plane current directions.
We consider two types of pairing symmetry in FeBS, either
$s_{\pm}$ or $s_{++}$ one.
In Sec. IV,  we present  numerically calculated results of phase dependencies of dc Josephson current for different orientations of the junctions.
We show that $c$-axis oriented junction can be used
to distinguish between the $s_{\pm}$ and the $s_{++}$-wave types of symmetry in FeBS.
We also present the temperature dependencies of the maximum Josephson current.
In Sec. V experimental data for PbIn/Ba$_{1-x}$K$_{x}$(FeAs)$_2$ point-contact Josephson junctions are presented which are consistent with theoretical predictions
for the  $s_{\pm}$ model.
We summarize the results and formulate conclusions in Sec. VI.

\section{\label{sec2}Tight-binding model for single-orbital case}

In this section, we formulate Green's function approach for calculation of dc Josephson current in single-orbital tight-binding models.
First, we consider  the procedure of calculation of 1D Josephson current for one-dimensional $S/I/S$ junctions,
where $S$ is conventional spin-singlet $s$-wave superconductor and $I$ is an insulating layer.
Then we outline the same procedure for $S/I/S_d$ junctions, where $S_d$ is a spin-singlet  single-orbital $d$-wave superconductor.
We demonstrate that our tight-binding Green's function approach reproduces the previous results for both $S/I/S$ and $S/I/S_d$ Josephson current.
In the end of this Section, we discuss an alternative plane wave approach
for the calculation of the Josephson current.

\subsection{Model of $S/I/S$ Josephson junction}

\begin{figure}[b]
\centerline{\includegraphics[width=8cm]{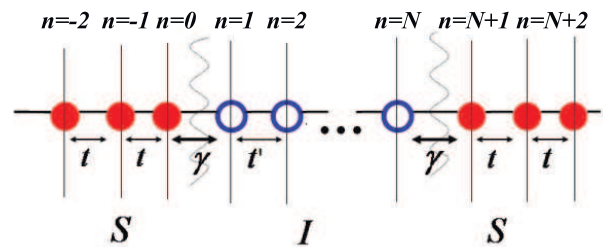}}
\caption{Schematic illustration of $1D$ model of the S/I/S Josephson junction. }
\label{SNS}
\end{figure}

We consider the 1D tight-binding model of $S/I/S$ Josephson junction as depicted in Fig.\ref{SNS}.
In the left and right parts of Fig.\ref{SNS}, red filled circles represent sites of
$s$-wave superconductor $S$ with hopping amplitude $t$.
In the middle of Fig.\ref{SNS}, there are $N$ sites of an insulator,
which we represent as blue circles with hopping $t'$ between them.
At the $S/I$ and $I/S$ boundaries, we choose the equal magnitude of
the hopping parameters $\gamma$ in Fig.\ref{SNS}.
We  assume that superconductors which form $S/I/S$ Josephson junction
are the same with common pair potential $\Delta_0$.
For simplicity, we assume that
the lattice spacing $a$ in $S$ and $I$ are the same and $a=1$.
To calculate the Josephson current across $S/I/S$ junction,
we must construct a Green's function of the whole  system.
The simplest way to do it is to construct the Green's functions in the $S,~I,~S$ regions first and then to match them at the boundaries.
We define temperature Green's functions in the tight-binding model in the following form:

\begin{equation}
\begin{aligned}
&G_{n,j}(\tau_1, \tau_2)=-\langle T_{\tau}c_{\uparrow}(n,\tau_1)c^{+}_{\uparrow}(j,\tau_2)\rangle,\\
&F_{n,j}(\tau_1, \tau_2)=\langle T_{\tau}c^{+}_{\downarrow}(n,\tau_1)c^{+}_{\uparrow}(j,\tau_2)\rangle,\\
&\tilde{G}_{n,j}(\tau_1, \tau_2)=-\langle T_{\tau}c^{+}_{\downarrow}(n,\tau_1)c_{\downarrow}(j,\tau_2)\rangle,\\
&\tilde{F}_{n,j}(\tau_1, \tau_2)=\langle T_{\tau}c_{\uparrow}(n,\tau_1)c_{\downarrow}(j,\tau_2)\rangle,
\end{aligned}
\end{equation}

\noindent with creation (annihilation) operator
$c_{\sigma}^+(n,\tau_i) \left(c_{\sigma}(n, \tau_i)\right)$  of an electron with spin $\sigma$ on $n$ site  and imaginary time ordering operator $T_{\tau}$.

After the differentiation of Green's functions with respect to $\tau_1$,
one can obtain the lattice version of  Gorkov's equations:
%
%
%

\begin{equation}
\left\{
\begin{aligned}
&(i\omega - \mu)G^{\omega}_{n,j}-\sum_{l}{t_{n,l}G^{\omega}_{l,j}}+\Delta_nF_{n,j}^{\omega}=\delta_{n,j},\\
&(i\omega + \mu)F^{\omega}_{n,j}+\sum_{l}{t_{n,l}F^{\omega}_{l,j}}+\Delta^{*}_nG_{n,j}^{\omega}=0,\\
&(i\omega - \mu)\tilde{F}^{\omega}_{n,j}-\sum_{l}{t_{n,l}\tilde{F}^{\omega}_{l,j}}+\Delta_n\tilde{G}_{n,j}^{\omega}=0,\\
&(i\omega + \mu)\tilde{G}^{\omega}_{n,j}+\sum_{l}{t_{n,l}\tilde{G}^{\omega}_{l,j}}+\Delta^{*}_n\tilde{F}_{n,j}^{\omega}=\delta_{n,j},
\end{aligned}
\right.
\label{Eq_Gr_s}
\end{equation}

\noindent where $t_{n,l}=t$ for $l=n\pm1$, $t_{n,l}=0$ for other values of $l$, $\omega=\pi T (2m+1)$ is the Matsubara frequency,
and $T$ is the temperature.
In the insulating region, we choose $\Delta_n=0$ in Eq. (\ref{Eq_Gr_s}).
One can find exact solutions of Eqs. (\ref{Eq_Gr_s}) as follows,
\begin{eqnarray}
\left(
\begin{array}{c}
G^{\omega, S_L}_{n,j} \\
F^{\omega, S_L}_{n,j}
\end{array}
\right)=a_{1j}\left(
\begin{array}{c}
\beta \\
e^{i\phi}
\end{array}
\right)e^{-ikn}
+
a_{2j}\left(
\begin{array}{c}
\beta^{-1} \\
e^{i\phi}
\end{array}
\right)e^{ikn}
\label{grl}
\end{eqnarray}

\noindent in the left superconductor,

\begin{eqnarray}
\left(
\begin{array}{c}
G^{\omega, S_R}_{n,j} \\
F^{\omega, S_R}_{n,j}
\end{array}
\right)=b_{1j}\left(
\begin{array}{c}
\beta \\
1
\end{array}
\right)e^{ikn}
+
b_{2j}\left(
\begin{array}{c}
1 \\
\beta
\end{array}
\right)e^{-ikn}
\label{grr}
\end{eqnarray}

\noindent in the right superconductor, and

\begin{equation}
\begin{aligned}
&\left(
\begin{array}{c}
G^{\omega, I}_{n,j} \\
F^{\omega, I}_{n,j}
\end{array}
\right)=c_{1j}\left(
\begin{array}{c}
1 \\
0
\end{array}
\right)e^{qn}
+
c_{2j}\left(
\begin{array}{c}
1 \\
0
\end{array}
\right)e^{-qn} \\&
+
c_{3j}\left(
\begin{array}{c}
0 \\
1
\end{array}
\right)e^{qn}
+
c_{4j}\left(
\begin{array}{c}
0 \\
1
\end{array}
\right)e^{-qn}
- \left(
\begin{array}{c}
1 \\
0
\end{array}
\right)\frac{e^{-q|n-j|}}{2t\sinh{q}}
\label{gri}
\end{aligned}
\end{equation}

\noindent in the insulator.
Here, $\beta=-i(\sqrt{\omega^2+|\Delta|^2}+\omega)/|\Delta|$, $\varphi=\varphi_R - \varphi_L$
 and $k$ ($q$)
are the phase difference between left and right superconductor
and momentum of quasiparticle in superconductor (insulator), respectively.
We assumed also, that in Eqs. (\ref{grl})-(\ref{gri}) the quasiclassical approximation ($\Delta \ll \mu,t,t'$) is applied.

Unknown coefficients $a_{1j}$, $a_{2j}$, $b_{1j}$, $b_{2j}$, $c_{1j}$, $c_{2j}$, $c_{3j}$, and $c_{4j}$ in Eqs. (\ref{grl}) - (\ref{gri})
can be obtained from matching the Green's functions (\ref{grl}) - (\ref{gri}) at the $S/I$ and $I/S$ interfaces.
The boundary conditions for multiorbital metals in tight-binding approximation have been proposed recently
\cite{bcrus,Burmistrova2013}.
For temperature Green's functions,
these boundary conditions in the quasiclassical approximation at $S/I$ boundary have the form:

\begin{equation}
\left\{
\begin{aligned}
&tG^{\omega,S_L}_{1,j}=\gamma G^{\omega,I}_{1,j},\\
&tF^{\omega,S_L}_{1,j}=\gamma F^{\omega,I}_{1,j},\\
&\gamma G^{\omega,S_L}_{0,j}=t'G^{\omega,I}_{0,j},\\
&\gamma F^{\omega,S_L}_{0,j}=t'F^{\omega,I}_{0,j},
\end{aligned}
\right.\label{bc1_left}
\end{equation}

and the following form at $I/S$ boundary:

\begin{equation}
\left\{
\begin{aligned}
&t'G^{\omega,I}_{N+1,j}=\gamma G^{\omega,S_R}_{N+1,j},\\
&t'F^{\omega,I}_{N+1,j}=\gamma F^{\omega,S_R}_{N+1,j},\\
&\gamma G^{\omega,I}_{N,j}=tG^{\omega,S_R}_{N,j},\\
&\gamma F^{\omega,I}_{N,j}=tF^{\omega,S_R}_{N,j}.
\end{aligned}
\right.\label{bc1_right}
\end{equation}

In the same way, one can find the other pair of Green's functions $\tilde{G}^{\omega}_{n,j}$ and $\tilde{F}^{\omega}_{n,j}$
from Eq. (\ref{Eq_Gr_s}).


The Josephson current across $1D$ $S/I/S$  junction in the tight-binding model
is given by the following expression:

\begin{equation}
\begin{aligned}
I(\varphi) = \frac{eTt}{i\hbar}\sum_{\omega}{(G^{\omega}_{j,j+1}-G^{\omega}_{j+1,j}
+\tilde{G}^{\omega}_{j,j+1}-\tilde{G}^{\omega}_{j+1,j})}.
\label{current}
\end{aligned}
\end{equation}

\noindent Eq. (\ref{current}) is the generalization for lattice model
version of the Josephson current in the framework of Green's function approach.

Using Eqs. (\ref{grl}) - (\ref{bc1_right}), it is possible to derive analytically, that previous results
\cite{Beenakker1991,Kulik1977,Kulik1978,amb,Furusaki1991,Bagwell1992}
for Josephson tunneling across $S/I/S$ constriction for equal hopping parameters in $S$ and $I$ with $t=t'$
are reproduced by the present tight-binding approach:

\begin{equation}
\begin{aligned}
I(\varphi) = \frac{e\Delta_0\sigma_N\sin\varphi}{2\sqrt{1-\sigma_N\sin^2(\frac{\varphi}{2})}}\tanh{\frac{\Delta_0\sqrt{1-\sigma_N\sin^2(\frac{\varphi}{2})}}{2T}},
\end{aligned}
\label{curphi}
\end{equation}

\noindent where $\sigma_N$ is the transparency of the $S/I/S$ junction in the normal state.
The transparency $\sigma_N$ is equal to unity in the case of the direct contact ($N=0$ layers of insulator atoms) with equal hopping parameters
in the bulk and at the interface, $\gamma=t$.

%
%
%
%
For the direct contact, the expression of $\sigma_N$ has the following form:

\begin{equation}
\begin{aligned}
\sigma_N = \frac{2\sigma_1^2(1-\cos{2k})}{\sigma_1^4-2\sigma_1^2\cos{2k}+1},
\end{aligned}
\label{tr1}
\end{equation}

\noindent with $\sigma_1=t^2/ \gamma^2$.
In the case of  $\gamma =t$ with nonzero length of an insulating region,
$i.e.$,  $N \neq 0$, transparency $\sigma_N$ of the $S/I/S$ junction has the following form:

\begin{equation}
\begin{aligned}
\sigma_N = \frac{4\sin^2{k}\sin^2{q}}{\Big(\sigma_2^2 + \sigma_3^2 -2\sigma_2\sigma_3\cos(2qN)\Big)^2},
\end{aligned}
\label{tr2}
\end{equation}

\noindent where $\sigma_2=1-\cos(k+q)$ and $\sigma_3=1-\cos(k+q)$.

Thus, based on
our tight binding Green's functions, Eq. (\ref{curphi}) reproduces
well-known previous results\cite{Beenakker1991,Kulik1977,Kulik1978,amb,Furusaki1991,Bagwell1992},
with generalized definition of the normal state transparency Eqs. (\ref{tr1}) and (\ref{tr2}).


\subsection{Model of  $S/I/S_d$ Josephson junction}

In this subsection, we extend the present tight-binding Green's functions approach to single orbital $d$-wave supercondutor($S_{d}$).
We consider $2D$ model of $S/I/S_d$ planar junction, where pair potential
in $d$-wave superconductor has the form
$\Delta=2\Delta_d(\cos k_x-\cos k_y)$ for zero misorientation angle and
$\Delta=4\Delta_d\sin k_x\sin k_y$ for $\pi/4$ misorientation angle.
We assume that the energy dispersion of
both left and right superconductors has the form $\varepsilon_N=2t(\cos k_x + \cos k_y) + \mu_N$
and that in an insulating region $\varepsilon_I=2t(\cos k_x + \cos k_y) + \mu_I$ with $\mu_I > \mu_N$.
In the actual numerical calcuation, Josepshon current is expressed by the
summation of all possible values of $k_y$.
With the increase of the thickness of the insulator $N$,
the quasiparticles around perpendicular injection to the insulator provide
dominant contributions to the total Josepshon current and the
contribution from the large values of $k_y$ are suppressed.

For zero misorientation angle,
surface Andreev bound states are absent.
For low transparent case, the qualitative feature of the
Josepshon current is similar to that of conventional $s$-wave superconductor.
However, for high transparent case,
current phase relation can deviate from  simple sinusoidal curret-phase relation proportional
reflecting on the $d$-wave symmetry, where $\phi$ denotes the macroscopic
phase difference between left and right superconductors.
Then free energy mimima can locate $\varphi=\pm \varphi_{0}$,
where $\varphi_{0}$ is neither 0 nor $\pm \pi$.
The reason can be understood if we decompose
Josephson current into components with
fixed $k_{y}$.
For the region with small values of $k_y$,
the obtained current phase relation is propotional to $\sin \varphi$.
On the other hand,
for the large magnitude of $k_{y}$,
Josepshon current is proportional to $-\sin \varphi$.
Then, after angle avrage of $k_{y}$,
first order term is relatively suppressed as compared to that of the second order term proporional to $\sin 2\varphi$.
Our calculations demostrate that increasing the length of an insulating region
upto $N=3$, the current phase relation becomes
that of $0$-junction.
In this case  contributions to the averaged Josephson current from the regions with large magnitude of $k_y$ are suppressed and that
from the  regions with small magnitude of $k_y$ prevail.
This feature obtained in the framework of our Green's function tight-binding  approach coincide qualitatively with the previous results derived in \cite{tanaka97} (Fig. 2 of [\onlinecite{tanaka97}]).

Next, we study the case with $\pi/4$ misorientation angle.
It is known that, the current phase relation becomes very unusual in this case.
The regions with positive and negative values of $k_y$ give rise to different phase dependencies of the Josephson current for fixed $k_y$ .
Positive values of $k_y$ correspond to $0$-junction and
negative one contiribute  to $\pi$-jucntion.
Then the first order term disappears.
Then, the free energy minima locates neither $0$ nor $\pm \pi$.

Our calculations demostrate that the above feature appears even
with increasing the number of an insulating layer up to $N=4$.
Current phase relation is proportional to $\sin 2\phi$ for
low transparent juntion with nonzero $N$.

These results obtained in the framework of the present lattice
Green's approach coincide qualitatively with the previous results derived in \cite{tanaka96,tanaka97} (Fig. 3 of [\onlinecite{tanaka97}]).

It is necessary to note that the same results as described above can be obtained not only in terms of Green's functions but also in terms of wave functions\cite{bag1}.
For this purpose one should solve Bogoliubov-de Gennes equations and find wave functions for a $s$-wave superconductor, an insulating region and a $d$-wave superconductor on the sites of the descrete lattice \cite{Burmistrova2013}.
However, calculations of the total Josephson current in terms of wave functions are inconvenient for the averaging of the Josephson current over all possible values of $k_y$  than in terms of Green's functions and lead to numerical errors.
Therefore, in the following sections we use the tight-binding Green's functions approach to obtain the averaged Josephson current in  FeBS junctions.

\section{model for the contact between s-wave superconductor and a FeBS}{\label{sec3}}

In this section we consider Josephson transport across the $S/I/S_p$ junctions, where $S$ is a single-orbital $s$-wave superconductor,
$I$ is an insulating layer and $S_p$ is a FeBS.
First, we consider the procedure of the calculation of 2D Josephson current for the (100) oriented $S/I/S_p$ junctions
for zero misorientation angle.
Then, we describe the same procedure for $S/I/S_p$ junctions along $c$-axis.

\begin{figure}[h]
\centerline{\includegraphics[width=8cm]{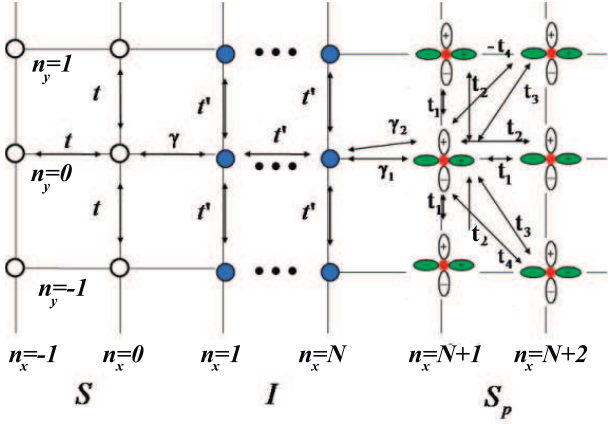}}
\caption{$2D$ tight-binding model of the (100) oriented $S/I/S_p$ junction. }
\label{SISp}
\end{figure}

\subsection{2D model of the $S/I/S_p$ Josephson junction with a (100) oriented FeBS}

\begin{figure}[h]
\centerline{\includegraphics[width=8cm]{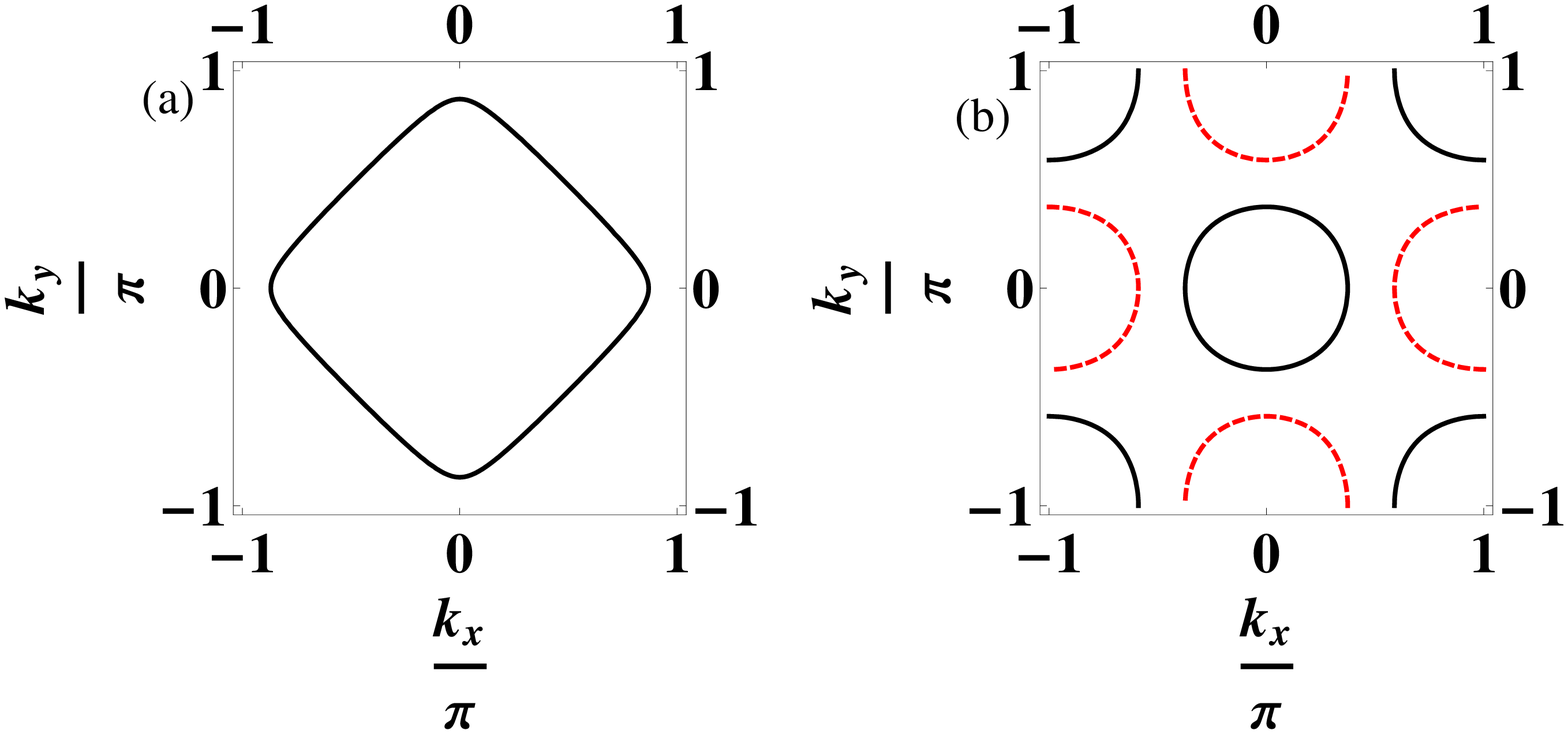}}
\caption{Fermi surfaces  of the $S/I/S_p$ junction with a (100) oriented FeBS. a) Fermi surface of a single-orbital s-wave superconductor, b) Fermi surface of a FeBS.}
\label{Fermi}
\end{figure}

In Fig. \ref{SISp} a two-dimensional crystallographic plane of a single-orbital $s$-wave superconductor $S$ (empty circles on left side of Fig. \ref{SISp}), $N$ atomic layers of an insulator (blue filled circles in the middle of Fig. \ref{SISp}) and a FeBS in the right part of Fig. \ref{SISp} are presented.
The minimal model to reproduce Fermi surfaces in a FeBS is a two-orbital model consists of $d_{xz}$ and $d_{yz}$ orbitals in iron \cite{rag}.
There are four hopping parameters $t_1$, $t_2$, $t_3$ and $t_4$ in this model, as shown in Fig. \ref{SISp}.
The Fermi surface of a FeBS in unfolded Brillouin zone is shown in Fig. \ref{Fermi}(b).

For the pair potentials,  the intra-orbital $s_{\pm}$
and  $s_{++}$ models  are  considered \cite{maz1,Kuroki,Kontani}.
The hopping between sites of a single-orbital superconductor $S$ and an insulator $I$ is described by parameters $t$ and $t'$, respectively.
The  hopping parameter across the interface between $S$ and $I$ is described by $\gamma$ and that between $I$ and $d_{xz}$ ($d_{yz}$)-orbitals in $S_p$
are described by $\gamma_1$ ($\gamma_2$). Due to the necessity to take into account at least two orbitals for correct description of the FeBS band structure, two hopping parameters $\gamma_1$  and $\gamma_2$ should be introduced in orbital space which describe an interface between single-band and two-band materials (instead of single hopping parameter $\gamma$ at the interface between two single-band materials \cite{Zhu1983}). The introduction of these two parameters provides a possibility to match coherently wave functions (Green functions) at this boundary and to describe the processes of interband scaterring microscopically, as it was demonstrated in \cite{bcrus,Burmistrova2013}
For simplicity, we assume that the lattice constants in $S,I$ and $S_p$ are equal.
To calculate  the Josephson current across $S/I/S_p$ junction we should construct Green's functions of the whole system.
The Green's function in $S,~I$ regions are presented in section \ref{sec2}A (Eqs. (\ref{grl}),(\ref{gri})).
\begin{widetext}
\begin{equation}
\begin{aligned}
&G_{\{n\},\{j\}}(\tau_1, \tau_2)=\begin{pmatrix} G^{\alpha\alpha}_{\{n\},\{j\}}(\tau_1, \tau_2) & G^{\alpha\beta}_{\{n\},\{j\}}(\tau_1, \tau_2)\\ G^{\beta\alpha}_{\{n\},\{j\}}(\tau_1, \tau_2) & G^{\beta\beta}_{\{n\},\{j\}}(\tau_1, \tau_2)\end{pmatrix}=\begin{pmatrix}
-\langle T_{\tau}c_{\uparrow}(\{n\},\tau_1)c^{+}_{\uparrow}(\{j\},\tau_2)\rangle & -\langle T_{\tau}c_{\uparrow}(\{n\},\tau_1)d^{+}_{\uparrow}(\{j\},\tau_2)\rangle\\
-\langle T_{\tau}d_{\uparrow}(\{n\},\tau_1)c^{+}_{\uparrow}(\{j\},\tau_2)\rangle & -\langle T_{\tau}d_{\uparrow}(\{n\},\tau_1)d^{+}_{\uparrow}(\{j\},\tau_2)\rangle
\end{pmatrix},
\\
&F_{\{n\},\{j\}}(\tau_1, \tau_2)=\begin{pmatrix} F^{\alpha\alpha}_{\{n\},\{j\}}(\tau_1, \tau_2) & F^{\alpha\beta}_{\{n\},\{j\}}(\tau_1, \tau_2)\\ F^{\beta\alpha}_{\{n\},\{j\}}(\tau_1, \tau_2) & F^{\beta\beta}_{\{n\},\{j\}}(\tau_1, \tau_2)\end{pmatrix}=\begin{pmatrix}
\langle T_{\tau}c^{+}_{\downarrow}(\{n\},\tau_1)c^{+}_{\uparrow}(\{j\},\tau_2)\rangle & \langle T_{\tau}c^{+}_{\downarrow}(\{n\},\tau_1)d^{+}_{\uparrow}(\{j\},\tau_2)\rangle\\
\langle T_{\tau}d^{+}_{\downarrow}(\{n\},\tau_1)c^{+}_{\uparrow}(\{j\},\tau_2)\rangle & \langle T_{\tau}d^{+}_{\downarrow}(\{n\},\tau_1)d^{+}_{\uparrow}(\{j\},\tau_2)\rangle
\end{pmatrix},
\\
&\tilde{G}_{\{n\},\{j\}}(\tau_1, \tau_2)=\begin{pmatrix} \tilde{G}^{\alpha\alpha}_{\{n\},\{j\}}(\tau_1, \tau_2) & \tilde{G}^{\alpha\beta}_{\{n\},\{j\}}(\tau_1, \tau_2)\\ \tilde{G}^{\beta\alpha}_{\{n\},\{j\}}(\tau_1, \tau_2) & \tilde{G}^{\beta\beta}_{\{n\},\{j\}}(\tau_1, \tau_2)\end{pmatrix}=\begin{pmatrix}
-\langle T_{\tau}c^{+}_{\downarrow}(\{n\},\tau_1)c_{\downarrow}(\{j\},\tau_2)\rangle & -\langle T_{\tau}c^{+}_{\downarrow}(\{n\},\tau_1)d_{\downarrow}(\{j\},\tau_2)\rangle\\
-\langle T_{\tau}d^{+}_{\downarrow}(\{n\},\tau_1)c_{\downarrow}(\{j\},\tau_2)\rangle & -\langle T_{\tau}d^{+}_{\downarrow}(\{n\},\tau_1)d_{\downarrow}(\{j\},\tau_2)\rangle
\end{pmatrix},
\\
&\tilde{F}_{\{n\},\{j\}}(\tau_1, \tau_2)=\begin{pmatrix} \tilde{F}^{\alpha\alpha}_{\{n\},\{j\}}(\tau_1, \tau_2) & \tilde{F}^{\alpha\beta}_{\{n\},\{j\}}(\tau_1, \tau_2)\\ \tilde{F}^{\beta\alpha}_{\{n\},\{j\}}(\tau_1, \tau_2) & \tilde{F}^{\beta\beta}_{\{n\},\{j\}}(\tau_1, \tau_2)\end{pmatrix}=\begin{pmatrix}
\langle T_{\tau}c_{\uparrow}(\{n\},\tau_1)c_{\downarrow}(\{j\},\tau_2)\rangle & \langle T_{\tau}c_{\uparrow}(\{n\},\tau_1)d_{\downarrow}(\{j\},\tau_2)\rangle\\
\langle T_{\tau}d_{\uparrow}(\{n\},\tau_1)c_{\downarrow}(\{j\},\tau_2)\rangle & \langle T_{\tau}d_{\uparrow}(\{n\},\tau_1)d_{\downarrow}(\{j\},\tau_2)\rangle
\end{pmatrix},
\end{aligned}
\label{Gpn}
\end{equation}
\end{widetext}

\noindent where
$c_{\sigma}^+(\{n\},\tau_i) \left(c_{\sigma}(\{n\}, \tau_i)\right)$ and
$d_{\sigma}^+(\{n\},\tau_i) \left(d_{\sigma}(\{n\}, \tau_i)\right)$
are creation (annihilation) operators
for the $d_{xz}$ and $d_{yz}$-orbitals with spin $\sigma$ at $\{n\}=(n_x,n_y)$, respectively.
$T_{\tau}$ is the imaginary time ordering operator.
Superscript $\alpha(\beta)$ corresponds to the $d_{xz}(d_{yz})$ orbital, respectively.
Differentiating the Green's functions (\ref{Gpn}) with respect to $\tau_1$,
expanding them in Fourier series and using Hamiltonian for $2D$ two-orbital model of a FeBS\cite{mor},
one can obtain the following Gorkov's equations:

\begin{widetext}
\begin{equation}
\left\{
\begin{aligned}
&(i\omega - \mu)G^{\alpha\alpha,\omega}_{\{n\},\{j\}}-\sum_{\{l\}}{t^{\alpha\alpha}_{\{n\},\{l\}}G^{\alpha\alpha,\omega}_{\{l\},\{j\}}}-\sum_{\{l\}}{t^{\alpha\beta}_{\{n\},\{l\}}G^{\alpha\beta,\omega}_{\{l\},\{j\}}}+\sum_{\{l\}}\Delta_{\{n\},\{l\}}F_{\{l\},\{j\}}^{\alpha\alpha,\omega}=\delta_{\{n\},\{j\}},\\
&(i\omega - \mu)G^{\alpha\beta,\omega}_{\{n\},\{j\}}-\sum_{\{l\}}{t^{\beta\beta}_{\{n\},\{l\}}G^{\alpha\beta,\omega}_{\{l\},\{j\}}}-\sum_{\{l\}}{t^{\beta\alpha}_{\{n\},\{l\}}G^{\alpha\alpha,\omega}_{\{l\},\{j\}}}+\sum_{\{l\}}\Delta_{\{n\},\{l\}}F_{\{l\},\{j\}}^{\alpha\beta,\omega}=0,\\
&(i\omega + \mu)F^{\alpha\alpha,\omega}_{\{n\},\{j\}}+\sum_{\{l\}}{t^{\alpha\alpha}_{\{n\},\{l\}}F^{\alpha\alpha,\omega}_{\{l\},\{j\}}}+\sum_{\{l\}}{t^{\alpha\beta}_{\{n\},\{l\}}F^{\alpha\beta,\omega}_{\{l\},\{j\}}}+\sum_{\{l\}}\Delta^*_{\{n\},\{l\}}G_{\{l\},\{j\}}^{\alpha\alpha,\omega}=0,\\
&(i\omega + \mu)F^{\alpha\beta,\omega}_{\{n\},\{j\}}+\sum_{\{l\}}{t^{\beta\beta}_{\{n\},\{l\}}F^{\alpha\beta,\omega}_{\{l\},\{j\}}}+\sum_{\{l\}}{t^{\beta\alpha}_{\{n\},\{l\}}F^{\alpha\alpha,\omega}_{\{l\},\{j\}}}+\sum_{\{l\}}\Delta^*_{\{n\},\{l\}}G_{\{l\},\{j\}}^{\alpha\beta,\omega}=0,\\
&(i\omega - \mu)G^{\beta\beta,\omega}_{\{n\},\{j\}}-\sum_{\{l\}}{t^{\beta\beta}_{\{n\},\{l\}}G^{\beta\beta,\omega}_{\{l\},\{j\}}}-\sum_{\{l\}}{t^{\beta\alpha}_{\{n\},\{l\}}G^{\beta\alpha,\omega}_{\{l\},\{j\}}}+\sum_{\{l\}}\Delta_{\{n\},\{l\}}F_{\{l\},\{j\}}^{\beta\beta,\omega}=\delta_{\{n\},\{j\}},\\
&(i\omega - \mu)G^{\beta\alpha,\omega}_{\{n\},\{j\}}-\sum_{\{l\}}{t^{\alpha\alpha}_{\{n\},\{l\}}G^{\beta\alpha,\omega}_{\{l\},\{j\}}}-\sum_{\{l\}}{t^{\alpha\beta}_{\{n\},\{l\}}G^{\beta\beta,\omega}_{\{l\},\{j\}}}+\sum_{\{l\}}\Delta_{\{n\},\{l\}}F_{\{l\},\{j\}}^{\beta\alpha,\omega}=0,\\
&(i\omega + \mu)F^{\beta\beta,\omega}_{\{n\},\{j\}}+\sum_{\{l\}}{t^{\beta\beta}_{\{n\},\{l\}}F^{\beta\beta,\omega}_{\{l\},\{j\}}}+\sum_{\{l\}}{t^{\beta\alpha}_{\{n\},\{l\}}F^{\beta\alpha,\omega}_{\{l\},\{j\}}}+\sum_{\{l\}}\Delta^*_{\{n\},\{l\}}G_{\{l\},\{j\}}^{\beta\beta,\omega}=0,\\
&(i\omega + \mu)F^{\beta\alpha,\omega}_{\{n\},\{j\}}+\sum_{\{l\}}{t^{\alpha\alpha}_{\{n\},\{l\}}F^{\beta\alpha,\omega}_{\{l\},\{j\}}}+\sum_{\{l\}}{t^{\alpha\beta}_{\{n\},\{l\}}F^{\beta\beta,\omega}_{\{l\},\{j\}}}+\sum_{\{l\}}\Delta^*_{\{n\},\{l\}}G_{\{l\},\{j\}}^{\beta\alpha,\omega}=0,
\end{aligned}
\right.\label{G_eq_p}
\end{equation}
\end{widetext}

\noindent Here, $t^{\alpha\alpha}_{\{n\},\{l\}}(t^{\beta\beta}_{\{n\},\{l\}})$ are the intra-orbital hopping parameters for $d_{xz}(d_{yz})$-orbital.
$t^{\alpha\beta}_{\{n\},\{l\}}(t^{\beta\alpha}_{\{n\},\{l\}})$ are the inter-orbital hopping parameters between the different orbitals, which have the following form:
\begin{widetext}
$t^{\alpha\alpha}_{\{n_x,n_y\},\{l_x,l_y\}}=t_1$ for $l_x=n_x\pm1,l_y=n_y$,

$t^{\alpha\alpha}_{\{n_x,n_y\},\{l_x,l_y\}}=t_2$ for $l_x=n_x,l_y=n_y\pm1$,

$t^{\alpha\alpha}_{\{n_x,n_y\},\{l_x,l_y\}}=t_3$ for $l_x=n_x\pm1,l_y=n_y\pm1$,

$t^{\alpha\alpha}_{\{n_x,n_y\},\{l_x,l_y\}}=0$ for the other conditions on the variables $l_x,n_x,l_y,n_y$;

$t^{\beta\beta}_{\{n_x,n_y\},\{l_x,l_y\}}=t_2$ for $l_x=n_x\pm1,l_y=n_y$,

$t^{\beta\beta}_{\{n_x,n_y\},\{l_x,l_y\}}=t_1$ for $l_x=n_x,l_y=n_y\pm1$,

$t^{\beta\beta}_{\{n_x,n_y\},\{l_x,l_y\}}=t_3$ for $l_x=n_x\pm1,l_y=n_y\pm1$,

$t^{\beta\beta}_{\{n_x,n_y\},\{l_x,l_y\}}=0$  for the other conditions on the variables $l_x,n_x,l_y,n_y$;

$t^{\alpha\beta}_{\{n_x,n_y\},\{l_x,l_y\}}=t^{\beta\alpha}_{\{n_x,n_y\},\{l_x,l_y\}}=t_4$ for $l_x=n_x\pm1,l_y=n_y\pm1$,

$t^{\alpha\beta}_{\{n_x,n_y\},\{l_x,l_y\}}=t^{\beta\alpha}_{\{n_x,n_y\},\{l_x,l_y\}}=0$ for the other conditions on the variables $l_x,n_x,l_y,n_y$.
\end{widetext}
In a similar way one can obtain the other Green's functions $\tilde{G}^{\alpha\alpha,\omega}_{\{n\},\{j\}},\tilde{G}^{\alpha\beta,\omega}_{\{n\},\{j\}},\tilde{G}^{\beta\alpha,\omega}_{\{n\},\{j\}},\tilde{G}^{\beta\beta,\omega}_{\{n\},\{j\}}$ and $\tilde{F}^{\alpha\alpha,\omega}_{\{n\},\{j\}},\tilde{F}^{\alpha\beta,\omega}_{\{n\},\{j\}},\tilde{F}^{\beta\alpha,\omega}_{\{n\},\{j\}}$ and $\tilde{F}^{\beta\beta,\omega}_{\{n\},\{j\}}$.

Placing the source terms $\delta_{\{n\},\{j\}}$ in Eqs. (\ref{Eq_Gr_s}),(\ref{G_eq_p})
into the insulating region $I$, one can see from Eqs. (\ref{G_eq_p}) that four upper and four lower equations(\ref{G_eq_p}) coincide, with $G^{\beta\alpha,\omega}_{\{n\},\{j\}}$
$G^{\beta\beta,\omega}_{\{n\},\{j\}}$,
$F^{\beta\alpha,\omega}_{\{n\},\{j\}}$,
and $F^{\beta\beta,\omega}_{\{n\},\{j\}}$
corresponding to $G^{\alpha\alpha,\omega}_{\{n\},\{j\}}$,
$G^{\alpha\beta,\omega}_{\{n\},\{j\}}$,
$F^{\alpha\alpha,\omega}_{\{n\},\{j\}}$
and $F^{\alpha\beta,\omega}_{\{n\},\{j\}}$, respectively.
Therefore, in order to calculate the Josephson current across this $S/I/S_p$ junction, it is enough to solve only either first four or last four equations in (\ref{G_eq_p}).

Solving first  four  Gorkov's equations (\ref{G_eq_p}), we obtain the Green's functions in the quasiclassical approximation ($\Delta_p << \mu,t_1,t_2,t_3,t_4$):

\begin{widetext}
\begin{eqnarray}&&
\left(
\begin{array}{c}
G^{\alpha\alpha,\omega}_{\{n\},\{j\}}\\
G^{\alpha\beta,\omega}_{\{n\},\{j\}}\\
F^{\alpha\alpha,\omega}_{\{n\},\{j\}}\\
F^{\alpha\beta,\omega}_{\{n\},\{j\}}
\end{array}
\right)=a_1
\left(
\begin{array}{c}
u_0(-k_{F_1})\\
v_0(-k_{F_1})\\
u_0(-k_{F_1})\beta^{(1)}_p(E,\Delta(-k_{F_1},k_y))\\
v_0(-k_{F_1})\beta^{(1)}_p(E,\Delta(-k_{F_1},k_y))
\end{array}
\right)e^{-ik_{F_1}n_x+ik_yn_y}+a_2\left(
\begin{array}{c}
u_0(k_{F_1})\\
v_0(k_{F_1})\\
u_0(k_{F_1})\tilde{\beta}^{(1)}_p(E,\Delta(k_{F_1},k_y))\\
v_0(k_{F_1})\tilde{\beta}^{(1)}_p(E,\Delta(k_{F_1},k_y))
\end{array}
\right)e^{ik_{F_1}n_x+ik_yn_y}\nonumber\\&&+a_3\left(
\begin{array}{c}
u_0(-k_{F_2})\\
v_0(-k_{F_2})\\
u_0(-k_{F_2})\beta^{(2)}_p(E,\Delta(-k_{F_2},k_y))\\
v_0(-k_{F_2})\beta^{(2)}_p(E,\Delta(-k_{F_2},k_y))
\end{array}
\right)e^{-ik_{F_2}n_x+ik_yn_y}+a_4\left(
\begin{array}{c}
u_0(k_{F_2})\\
v_0(k_{F_2})\\
u_0(k_{F_2})\tilde{\beta}^{(2)}_p(E,\Delta(k_{F_2},k_y))\\
v_0(k_{F_2})\tilde{\beta}^{(2)}_p(E,\Delta(k_{F_2},k_y))
\end{array}
\right)e^{ik_{F_2}n_x+ik_yn_y},
\label{GSp}
\end{eqnarray}
\end{widetext}

\noindent where


\begin{eqnarray}
\left(
\begin{array}{c}
u_0(k_x, k_y) \\
v_0(k_x, k_y)
\end{array}
\right)=\left(
\begin{array}{c}
1 \\
-\xi_{xx}(k_{F_i})/\xi_{xy}(k_{F_i})
\end{array}
\right)
\label{koeff}
\end{eqnarray}

\noindent and

\begin{equation}
\begin{aligned}
&\beta^{1(2)}_p = ie^{i\varphi}\frac{|\Delta_p(-k_{F_{1(2)}},k_y)|}{\sqrt{\omega^2+|\Delta_p(-k_{F_{1(2)}},k_y)|^2}+\omega},\\ &\tilde{\beta}^{1(2)}_p = -ie^{i\varphi}\frac{|\Delta_p(k_{F_{1(2)}},k_y)|}{\sqrt{\omega^2+|\Delta_p(k_{F_{1(2)}},k_y)|^2}-\omega}.
\label{beta}
\end{aligned}
\end{equation}

\noindent Here $\xi_{xx}=2t_1\cos(k_x)+2t_2\cos(k_y)+\mu$ and $\xi_{xy}=4t_4\sin(k_x)\sin(k_y)$ are dispersion relation of the $d_{xz}$ orbital and hybridization term, respectively, $\mu$ is a chemical potential and
$k_{F_{1(2)}}$ is momentum within the first (second) band in a FeBS.
In the similar way one can obtain the expressions for the Green's functions $\tilde{G}^{\alpha\alpha,\omega}_{\{n\},\{j\}},\tilde{G}^{\alpha\beta,\omega}_{\{n\},\{j\}}$ and $\tilde{F}^{\alpha\alpha,\omega}_{\{n\},\{j\}},\tilde{F}^{\alpha\beta,\omega}_{\{n\},\{j\}}$.

To build the Green's function of whole $S/I/S_p$ junction
one should match Green's functions in $S$, $I$ and $S_p$ regions (Eqs. (\ref{grl}),(\ref{gri}),(\ref{GSp}))
at both $S/I$ and $I/S_p$ interfaces.
The boundary conditions for the Green's functions in the tight-binding approximation can be found in a similar way as in  \cite{bcrus,Burmistrova2013}.
Due to the translational invariance of the structure in the direction parallel to the interface,
$k_y$ component of the momentum is conserved. Further, due to the
translational invariance of the considered structure the subscript with index $(y)$ corresponding to the coordinate along the boundary is omitted.
Thus, boundary conditions at the $S/I$ boundary have the form given in Eq. (\ref{bc1_left}).
At the $I/S_p$ interface
the boundary conditions have the following form \cite{bcrus,Burmistrova2013}:

%
%
\begin{equation}
\left\{
\begin{aligned}
&t_1G^{\alpha\alpha,\omega}_{N,j}+ 2t_3\cos{k_y}G^{\alpha\alpha,\omega}_{N,j} + 2it_4\sin{k_y}G^{\alpha\beta,\omega}_{N,j}\\&=\gamma_1 G^{\omega,I}_{N,j},\\
&t_1F^{\alpha\alpha,\omega}_{N,j}+ 2t_3\cos{k_y}F^{\alpha\alpha,\omega}_{N,j} + 2it_4\sin{k_y}F^{\alpha\beta,\omega}_{N,j}\\&=\gamma_1 F^{\omega,I}_{N,j},\\
&t_2G^{\alpha\beta,\omega}_{N,j}+ 2t_3\cos{k_y}G^{\alpha\beta,\omega}_{N,j} + 2it_4\sin{k_y}G^{\alpha\alpha,\omega}_{N,j}\\&=\gamma_2 G^{\omega,I}_{N,j},\\
&t_2F^{\alpha\beta,\omega}_{N,j}+ 2t_3\cos{k_y}F^{\alpha\beta,\omega}_{N,j} + 2it_4\sin{k_y}F^{\alpha\alpha,\omega}_{N,j}\\&=\gamma_2 F^{\omega,I}_{N,j},\\
&\gamma_1 G^{\alpha\alpha,\omega}_{N+1,j}+\gamma_2 G^{\alpha\beta,\omega}_{N+1,j}=t'G^{\omega,I}_{N+1,j},\\
&\gamma_1 F^{\alpha\alpha,\omega}_{N+1,j}+\gamma_2 F^{\alpha\beta,\omega}_{N+1,j}=t'F^{\omega,I}_{N+1,j}.
\end{aligned}
\right.\label{bc2}
\end{equation}

General expression for the Josephson current has the form
\begin{equation}
\begin{aligned}
I = \frac{eTt{^\prime}L{^\prime}}{2i\pi\hbar}&\int\sum_{\omega}(G^I_{j,j+1}-G^I_{j+1,j}+ \widetilde{G}^I_{j,j+1}-\widetilde{G}^I_{j+1,j})dk_y,
\end{aligned}
\label{current2}
\end{equation}
where $t{^\prime}$ is the hopping parameter inside the insulating region (see Fig.2) and $L{^\prime}=L/a$, $L$ is the width of the junction.

\subsection{3D model of the $S/I/S_p$ Josephson junction along $c$-axis
of FeBS}

Now we consider Josephson current across $S/I/S_p$ junction parallel
to $c$-axis of FeBS.

\begin{figure}[h]
\centerline{\includegraphics[width=8cm]{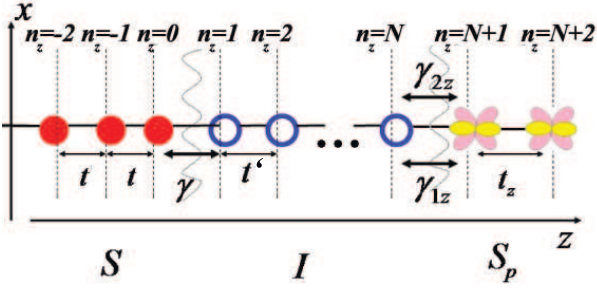}}
\caption{3D tight-binding model for $S/I/S_p$ junction along $z$-axis.
$t'$, $t$ and $t_z$ are hopping integrals along $z$-axis in $S$, $I$ and $S_p$, respectively.
$\gamma$ is hopping integral at the $S/I$ boundary.
$\gamma_{1z}$ and $\gamma_{2z}$ are hopping parameters between $I$ and $S_p$ for $xz$ and $yz$-orbitals, respectively}
\label{SISz}
\end{figure}

In Fig. \ref{SISz}, a single-orbital $s$-wave superconductor $S$
/insulator (I) / FeBS($S_{p}$) junction along $z$-direction is shown.
For 3D tight-binding model of FeBS, the hopping parameter $t_z$
between the same orbitals on the nearest neighbor sites in $z$-direction should be taken into account in addition to the hopping parameters $t_1,t_2,t_3,t_4$ in the $x-y$ plane. The existence of this hopping $t_z$ leads to light warping of cylindrical Fermi surface sheets in z-direction.
The main property of excitation spectrum of a FeBS as a function of $k_z$ is that for each fixed value of $k_{||}=(k_x,k_y)$ only one band crosses the Fermi level.
This means that for each value of $k_{||}$ only one of the bands contributes to the electronic transport.
In Fig. \ref{SISz}, $\gamma$, $\gamma_{1z}$ and $\gamma_{2z}$ are
hopping parameters across the $S/I$ and $I/S_p$ interfaces respectively.


For calculation of the Josephson current across $S/I/S_p$ junction along
$z$-axis of a FeBS one can define the temperature Green's function of a FeBS in the same way as in Eq. (\ref{Gpn}) . One can obtain the same set of Gorkov's equations like  (\ref{G_eq_p}) as in the previously considered case of the Josephson transport in $x-y$ plane of a FeBS, but with different definition of the hopping parameters: $t^{11}_{\{n\},\{l\}},t^{22}_{\{n\},\{l\}},t^{12}_{\{n\},\{j\}},t^{21}_{\{n\},\{j\}}$:
\begin{widetext}
$t^{(11)}_{\{n_x,n_y,n_z\},\{l_x,l_y,l_z\}}=t_1$ for $l_x=n_x\pm1,l_y=n_y,l_z=n_z$,

$t^{(11)}_{\{n_x,n_y,n_z\},\{l_x,l_y,l_z\}}=t_2$ for $l_x=n_x,l_y=n_y\pm1,l_z=n_z$,

$t^{(11)}_{\{n_x,n_y,n_z\},\{l_x,l_y,l_z\}}=t_3$ for $l_x=n_x\pm1,l_y=n_y\pm1,l_z=n_z$,

$t^{(11)}_{\{n_x,n_y,n_z\},\{l_x,l_y,l_z\}}=t_z$ for $l_x=n_x,l_y=n_y,l_z=n_z\pm1$,

$t^{(11)}_{\{n_x,n_y,n_z\},\{l_x,l_y\}}=0$ for the other conditions on the variables $l_x,n_x,l_y,n_y,l_z,n_z$;

$t^{(22)}_{\{n_x,n_y,n_z\},\{l_x,l_y,l_z\}}=t_2$ for $l_x=n_x\pm1,l_y=n_y,l_z=n_z$,

$t^{(22)}_{\{n_x,n_y,n_z\},\{l_x,l_y,l_z\}}=t_1$ for $l_x=n_x,l_y=n_y\pm1,l_z=n_z$,

$t^{(22)}_{\{n_x,n_y,n_z\},\{l_x,l_y,l_z\}}=t_3$ for $l_x=n_x\pm1,l_y=n_y\pm1,l_z=n_z$,

$t^{(22)}_{\{n_x,n_y,n_z\},\{l_x,l_y,l_z\}}=t_z$ for $l_x=n_x,l_y=n_y,l_z=n_z\pm1$,

$t^{(11)}_{\{n_x,n_y,n_z\},\{l_x,l_y\}}=0$ for the other conditions on the variables $l_x,n_x,l_y,n_y,l_z,n_z$;

$t^{(12)}_{\{n_x,n_y,n_z\},\{l_x,l_y,l_z\}}=t^{(21)}_{\{n_x,n_y\},\{l_x,l_y\}}=t_4$ for $l_x=n_x\pm1,l_y=n_y\pm1,l_z=n_z$,

$t^{(12)}_{\{n_x,n_y,n_z\},\{l_x,l_y,l_z\}}=t^{(21)}_{\{n_x,n_y\},\{l_x,l_y\}}=0$ for the other conditions on the variables $l_x,n_x,l_y,n_y,l_z,n_z$.
\end{widetext}

Solving Gorkov's equations for this 3D model of a FeBS one can obtain the Green's function in $S_p$ region, which has the same form as Eq.(\ref{GSp})  in the case of 2D model of a FeBS, but with another definition of the dispersion relation of the $d_{xz}$ orbital $\xi_{xx}$ in Eq.(\ref{koeff}): $\xi_{xx}=2t_1\cos(k_x)+2t_2\cos(k_y)+2t_z\cos(k_z)+\mu$. As a result, one can obtain the expressions for components of  Green's functions $\tilde{G}^{\alpha\alpha,\omega}_{\{n\},\{j\}},\tilde{G}^{\alpha\beta,\omega}_{\{n\},\{j\}},\tilde{G}^{\beta\alpha,\omega}_{\{n\},\{j\}},\tilde{G}^{\beta\beta,\omega}_{\{n\},\{j\}}$ and $\tilde{F}^{\alpha\alpha,\omega}_{\{n\},\{j\}},\tilde{F}^{\alpha\beta,\omega}_{\{n\},\{j\}},\tilde{F}^{\beta\alpha,\omega}_{\{n\},\{j\}},\tilde{F}^{\beta\beta,\omega}_{\{n\},\{j\}}$ for 3D model of a FeBS.  Green's functions for $S$ and $I$ regions can be found in a
similar way as in section \ref{sec2}A.

The boundary conditions for  Green's functions in the tight-binding approximation for  transport along $z$-axes can be found in the similar way \cite{bcrus,Burmistrova2013} as in section \ref{sec3}A and they have a simpler form than in the case of transport in $x-y$ plane. Due to the translational invariance of the structure in the direction parallel to the interface,
$k_{||}=(k_x,k_y)$ component of the momentum is conserved. Further, due to the
translational invariance of considered structure the subscripts with indices $(x,y)$ corresponding to the coordinate of an atom in a direction parallel to the boundary is omitted.
Thus, boundary conditions at the $S/I$ boundary coincide with Eq. (\ref{bc1_left}).
For the $I/S_p$ boundary we obtain the following boundary conditions in $z$-direction  in the quasiclassical approximation ($\Delta_0,\Delta_p,\Delta_p' << \mu_I,\mu,\mu_N,t,t',t_1,t_2,t_3,t_4,t_z$) \cite{bcrus,Burmistrova2013}:

\begin{equation}
\left\{
\begin{aligned}
&t_zG^{\alpha\alpha}_{N+1,j}=\gamma_{1z} G^{I}_{N+1,j},\\
&t_zF^{\alpha\alpha}_{1,j}=\gamma_{1z} F^{I}_{N+1,j},\\
&t_zG^{\alpha\beta}_{N+1,j}=\gamma_{2z} G^{I}_{N+1,j},\\
&t_zF^{\alpha\beta}_{N+1,j}=\gamma_{2z} F^{I}_{N+1,j},\\
&\gamma_{1z}G^{\alpha\alpha}_{N,j}+\gamma_{2z}G^{\alpha\beta}_{N,j}=tG^{I}_{N,j},\\
&\gamma_{1z}F^{\alpha\alpha}_{N,j}+\gamma_{2z}F^{\alpha\beta}_{N,j}=tF^{I}_{N,j}.
\end{aligned}
\right.\label{bc3}
\end{equation}


The Josephson current across $S/I/S_p$ junction is described by the sum  over all possible values of $k_{||}$ of  Eq. \ref{current2}, where $k_y$ should be replaced by $k_{||}=(k_x,k_y)$.

\section{Numerical results}{\label{sec4}}

In this section we present the results of numerical calculations of  the Josephson current across $S/I/S_p$ junction.
We calculate the averaged Josephson current by summing all possible  $k_{||}$   for two
models of pairing symmetry in a FeBs: the $s_{\pm}$ model with order parameter $\Delta=4\Delta_p\cos k_x\cos k_y$ with $\Delta_p=0.008$ (eV)
and the $s_{++}$ model with order parameter $\Delta=2\Delta_p(\cos k_x+\cos k_y)+\Delta_p'$ with $\Delta_p=0.001, \Delta_p'=0.0042$ (eV).
We choose $\Delta_0=0.002(eV)$ as the pair potential in $S$ .

There is a number of factors which influence the Josephson current averaged over $k_{||}$.

\begin{figure}
\begin{center}
    \begin{tabular}{cc}
      \resizebox{43mm}{!}{\includegraphics{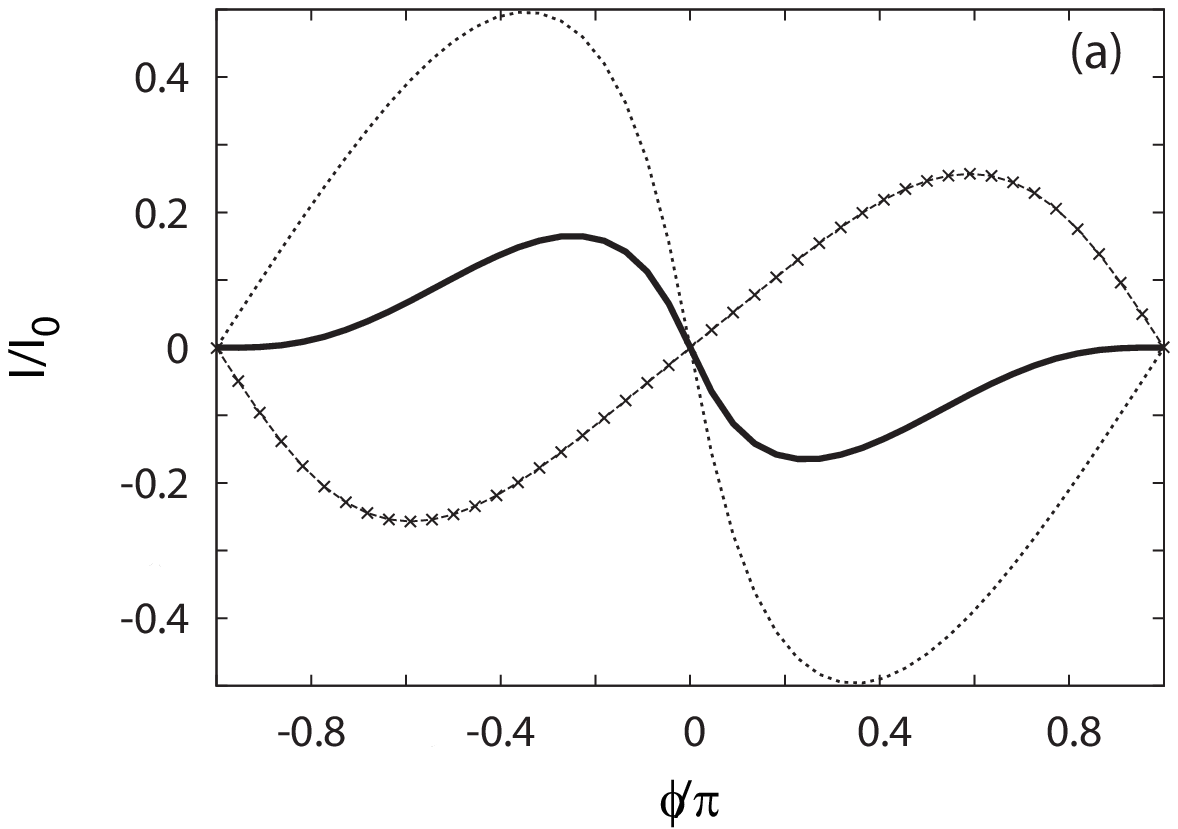}} &
      \resizebox{43mm}{!}{\includegraphics{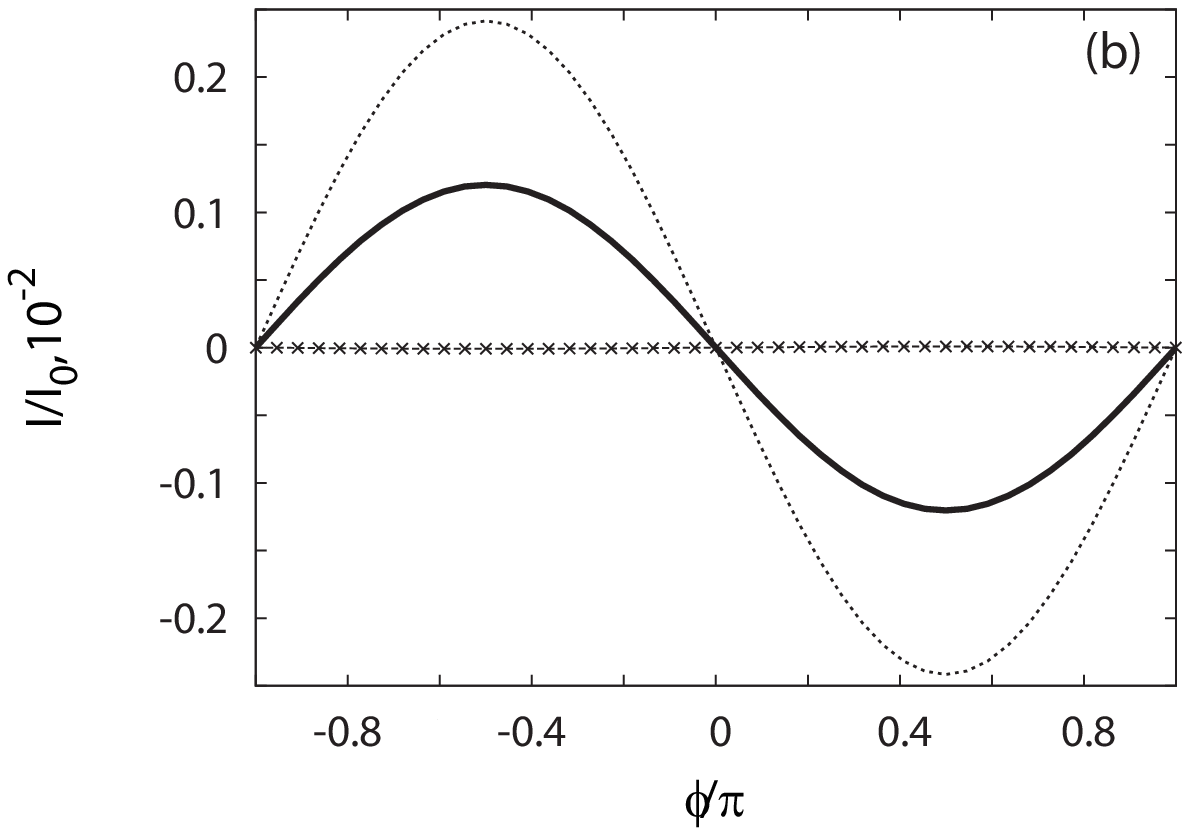}} \\
    \end{tabular}
\caption{The CPR for the (100) oriented $S/I/S_p$ junction for $\gamma_1=0.02, \gamma_2=0.2$ and $T/T^s_c \approx 0.02$.
The solid line corresponds to the total Josephson current, dotted line corresponds to the contribution from $|k_y|<\pi/2$. The line with crosses shows the contribution from $|k_y|>\pi/2$ .\textbf{ $I_0=e\Delta_0L^{\prime}/2\pi\hbar$. }Fig. \ref{R1}a corresponds to the direct contact, Fig. \ref{R1}b  corresponds to $N=3$  atomic layers in the insulating region.}
    \label{R1}
  \end{center}
  \end{figure}

 \begin{figure}
\begin{center}
    \begin{tabular}{cc}
      \resizebox{43mm}{!}{\includegraphics{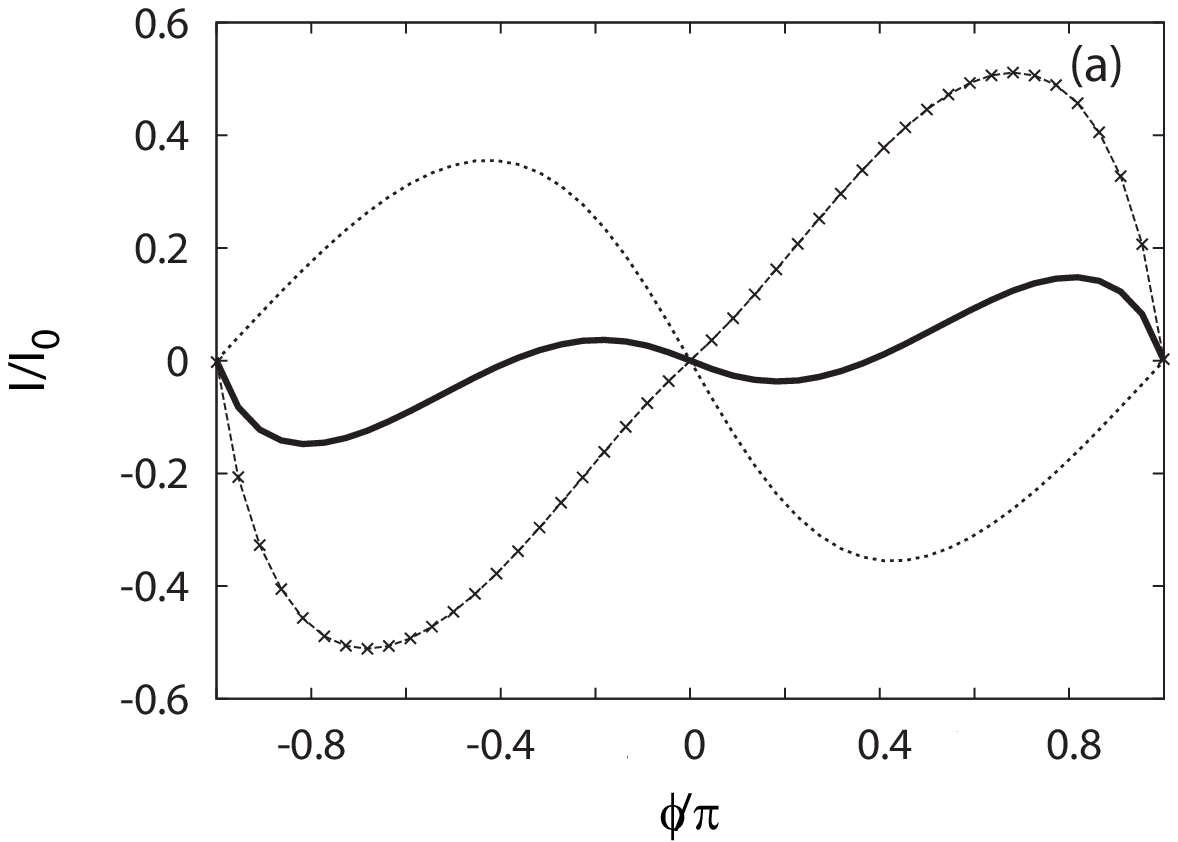}} &
      \resizebox{43mm}{!}{\includegraphics{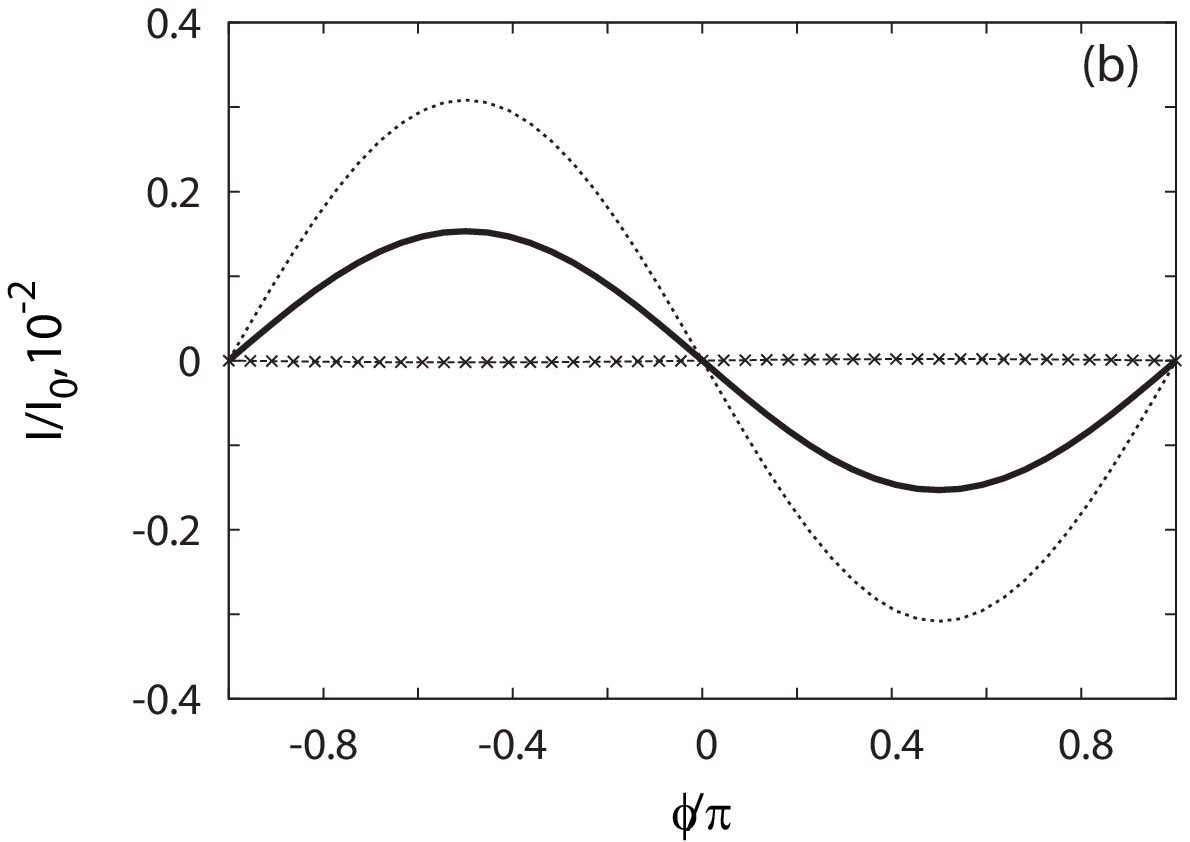}} \\
    \end{tabular}
\caption{ The same as in Fig.\ref{R1}, but $\gamma_1=0.02, \gamma_2=0.3$.}
    \label{R2}
  \end{center}
  \end{figure}

 \begin{figure}
\begin{center}
    \begin{tabular}{cc}
      \resizebox{43mm}{!}{\includegraphics{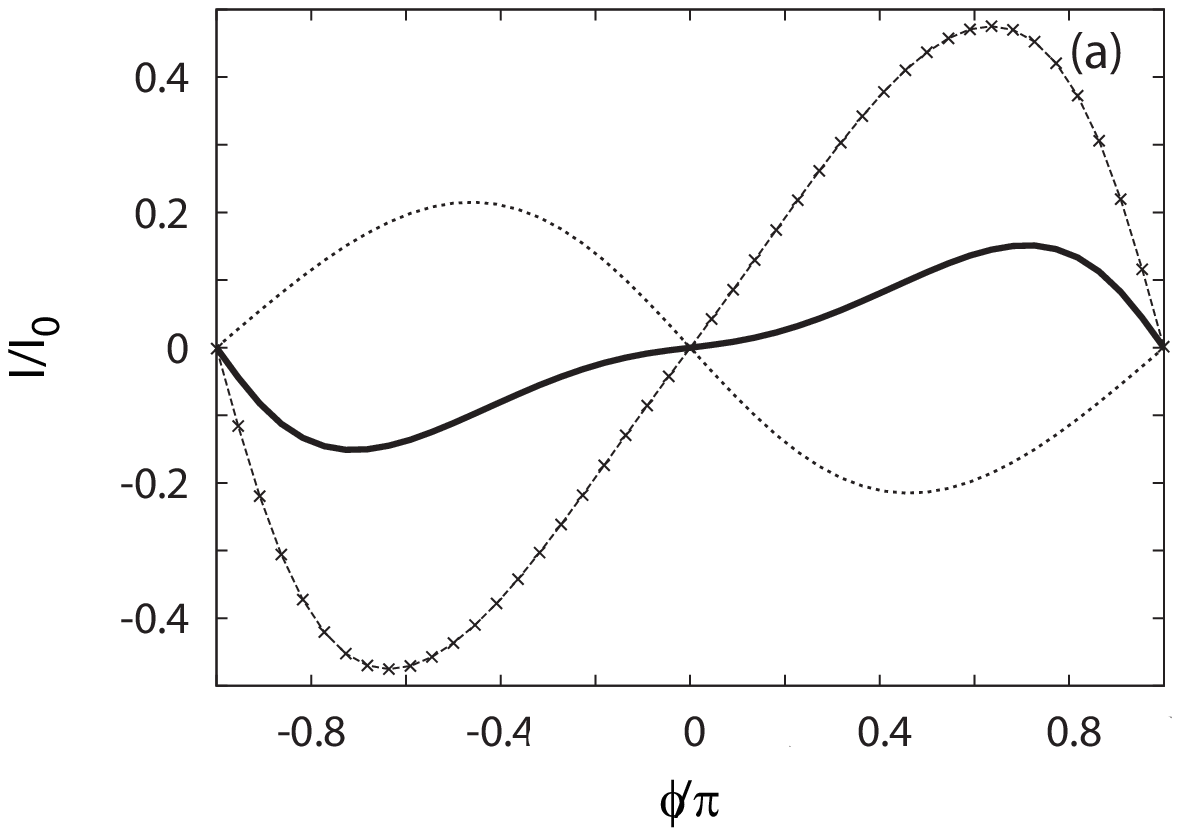}} &
      \resizebox{43mm}{!}{\includegraphics{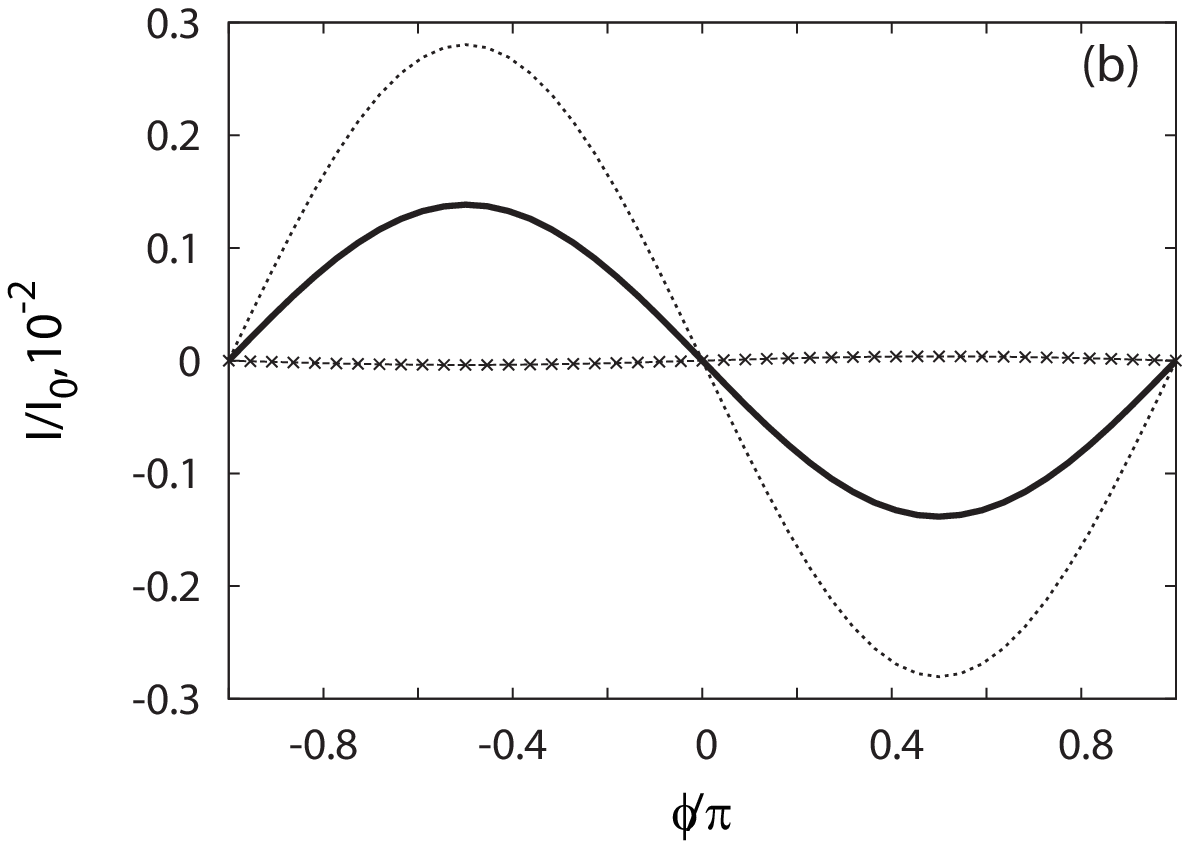}} \\
    \end{tabular}
\caption{ The same as in Fig.\ref{R1}, but $\gamma_1=0.02, \gamma_2=0.4$.}
    \label{R3}
  \end{center}
  \end{figure}

 \begin{figure}
\begin{center}
    \begin{tabular}{cc}
      \resizebox{43mm}{!}{\includegraphics{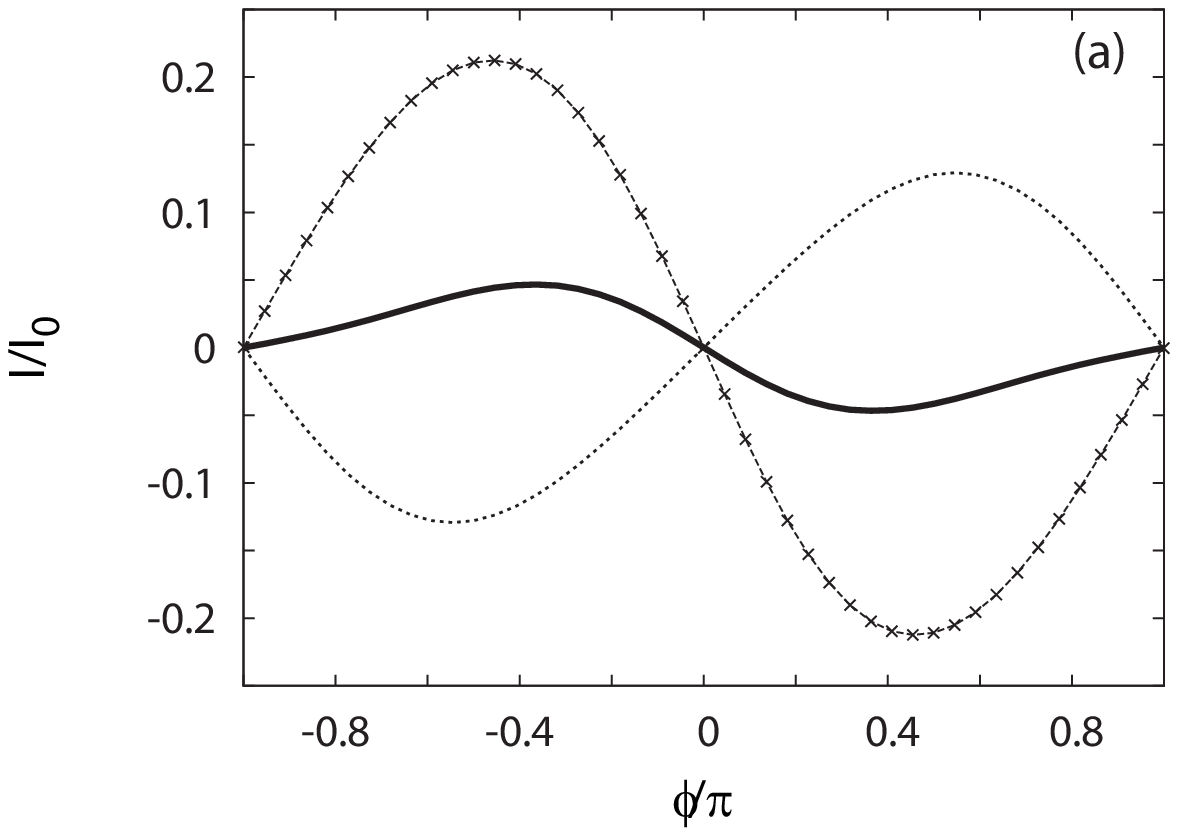}} &
      \resizebox{43mm}{!}{\includegraphics{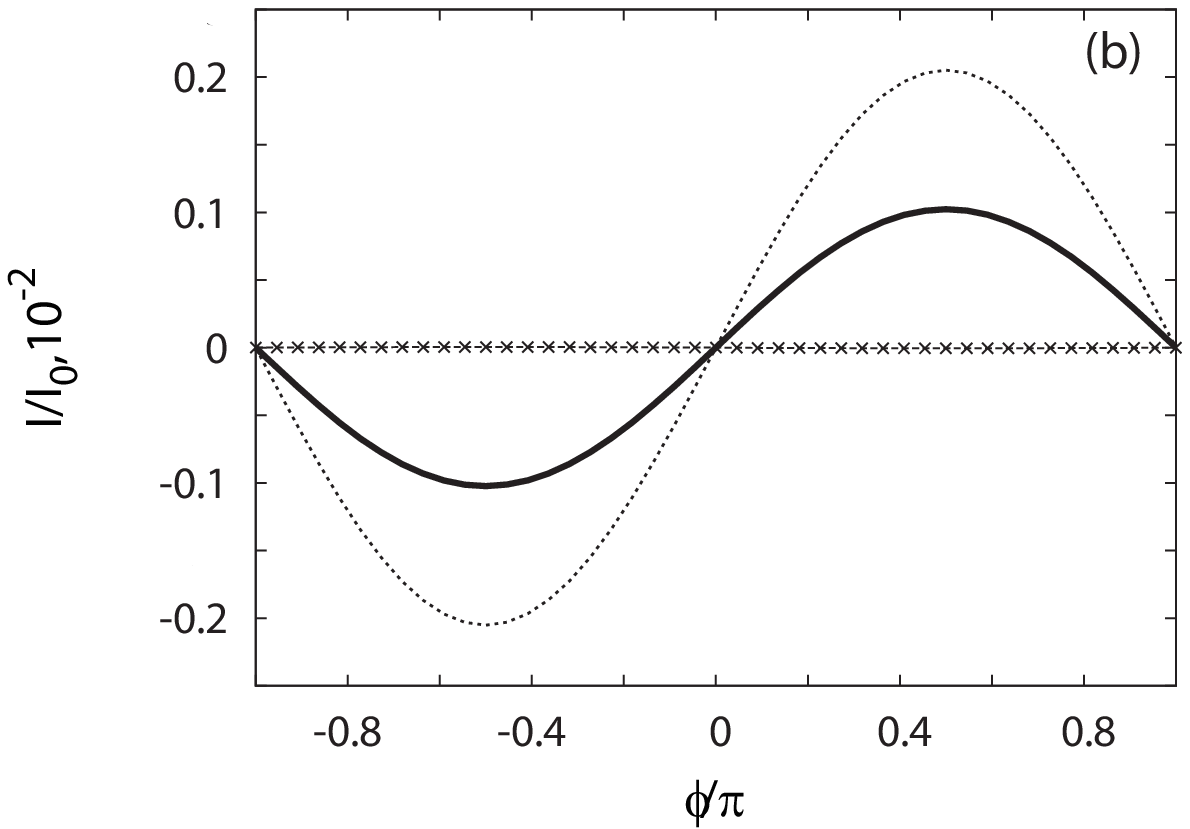}} \\
    \end{tabular}
\caption{ The same as in Fig.\ref{R1}, but $\gamma_1=0.2, \gamma_2=0.02$.}
    \label{R4}
  \end{center}
  \end{figure}

1. Sensitivity of Josephson current to the values of hopping parameters at the $I/S_p$ interface $\gamma_1$ and $\gamma_2$.

2. Influence of a Fermi surface size of an s-wave superconductor: the Josephson current strongly depends on the values of $k_{||}$, therefore, variation of the size of the Fermi surface in S leads to changes of relative contributions to the averaged Josephson current from regions with different $k_{||}$.

3. Influence of the length of an insulating layer: increasing this length leads to the suppression of the contributions from large $k_{||}$ to the averaged Josephson current.


\subsection{Current-phase relation in $S/I/S_p$ junctions}

First we present the results of  numerical calculations of current-phase relation (CPR) in (100) oriented $S/I/S_p$ Josephson junctions, when charge transport occurs in $x-y$ planes of a FeBS.
We choose the normal excitation spectrum in $S$ in the form $\varepsilon_N=2t(\cos k_x + \cos k_y) + \mu_N$, where $t=-0.3$ and $\mu_N=0.05$ in order to provide large size of the Fermi surface in S. Consequently, areas with large $k_y$ in a FeBS contribute to the current (Fig. \ref{Fermi}).
We use the following values of the hopping parameters and chemical potential in a FeBS:
$t_1 = -0.1051$, $t_2 = 0.1472$, $t_3 = -0.1909$, $t_4 = -0.0874$ and $\mu=-0.081$ (eV), according to Ref. \onlinecite{mor}, and suppose relatively low temperature $T/T^s_c \approx 0.02$, where $T^s_c$ is the critical temperature of the conventional $s$-wave superconductor.
In the insulating region, we choose the normal excitation spectrum in the form of $\varepsilon_I=2t'(\cos k_x + \cos k_y) + \mu_I$ with hopping parameter $t'=-0.3$ (eV) and chemical potential $\mu_I=1.2$ (eV).

The CPR for a direct $S/I/S_p$ contact is depicted in Fig.\ref{R1},a for hopping parameters across this boundary $\gamma_1=0.02, \gamma_2=0.2$.  Here the solid line corresponds to total Josephson current, the dotted line corresponds to the Josephson current  averaged over $k_y<\pi/2$ (over the values of $k_y$ belonging to the hole and electron pockets of a FeBS near $(k_x,k_y) = (0,0)$ and $(\pm \pi,0)$, respectively (Fig. \ref{Fermi},a)), while the line with crosses corresponds to the Josephson current  averaged over $k_y>\pi/2$ (the values of $k_y$ belonging to the electron and hole pockets of a FeBS near $(k_x,k_y) = (0,\pi)$ and $(\pm \pi,\pi$) (Fig. \ref{Fermi},a)). The contributions to the Josephson current from small $k_y$ contribute to the $\pi$-coupling, while the contributions from large values of $k_y$  to 0-contact. However, the sum of these two contributions leads to the formation of a $\pi$-contact. An increase of the length of an insulating barrier up to $N=3$ atomic layers leads to the suppression of the contributions to the averaged  Josephson current from large values of $k_y$, therefore the contribution from small values of $k_y$ dominates and the contact remains in the $\pi$-state for finite length of an insulator (Fig. \ref{R1},b).

Fig.\ref{R2},a shows the CPR averaged over $k_y$  for a direct $S/I/S_p$ contact for another parameter set $\gamma_2$ ($\gamma_1=0.02, \gamma_2=0.3$). For these values of hopping parameters the CPR is characterized by a stable equilibrium phase $0<\phi<\pi$, i.e. $\phi$-contact is realized. Increasing the length of an insulator up to $N=3$ leads to  suppression of the contributions to the averaged Josephson current from regions with large values of $k_y$ and the $\pi$-state is realized (Fig. \ref{R2},b).

The CPR averaged over $k_y$ for a direct $S/I/S_p$ contact is depicted in Fig. \ref{R3},a for $\gamma_1=0.02, \gamma_2=0.4$. In this case, the contribution from large $k_y$ values to the total averaged Josephson current prevails, hence the CPR is characterized by the stable equilibrium state at phase difference $\phi=0$ ($0$-contact). However, an increase of the length of an insulator up to $N=3$ atomic layers leads to the suppression of the large $k_y$ contributions to the averaged current and to the transition to a $\pi$-state (Fig. \ref{R3},b).

Finally, for the parameter set $\gamma_1=0.2, \gamma_2=0.02$ the CPR of the direct $S/I/S_p$ contact is shown in Fig. \ref{R4}a. In this case, the opposite situation in comparison with the previous cases (Figs. \ref{R1}-\ref{R3}) is realized, since here the contributions to the averaged current from small $k_y$ lead to appearance of $0$-contact, while the contributions from large $k_y$ - to $\pi$-contact.
 However for $N=0$,  as in the case shown in (Fig. \ref{R1},a), the sum of these contributions leads to the appearance of the resulting $\pi$-contact, because the contribution
 from regions with large values of $k_y$ dominates over that with small $k_y$. With an increase of the length of an insulator up to $N=3$ layers,
the contribution from large $k_y$ is suppressed and the junction goes in $0$-state (Fig.\ref{R4},b).


 \begin{figure}
\begin{center}
    \begin{tabular}{cc}
      \resizebox{106mm}{!}{\includegraphics{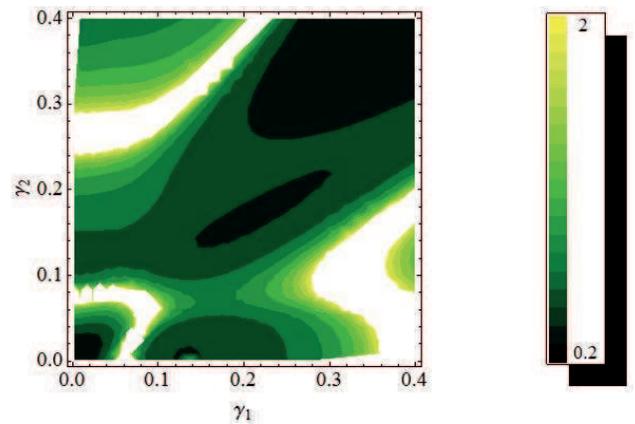}} &
 \\
    \end{tabular}
\caption{The ratio of the second and the first harmonics of the CPR for the (100) oriented $S/I/S_p$ Josephson junction for the the direct contact case
 as a function of hopping parameters across the boundary.}
    \label{Harmonic}
  \end{center}
  \end{figure}

In the case of $s_{++}$ pairing symmetry in a FeBS the order parameter has equal signs on each Fermi surface pocket (Fig. \ref{Fermi},a). Hence, for each value of $k_y$ and for any set of hopping amplitudes across the $I/S_p$ interface $\gamma_1$ and $\gamma_2$ we always obtain 0-contact. So, after averaging over all possible values of $k_y$ this $S/I/S_p$ junction has equilibrium phase which is equal to zero. Increasing the length of an insulating layer leads to the suppression of the contributions to the averaged current from large values of $k_y$, but the junction still remains in $0$-state.

 \begin{figure}
\begin{center}
    \begin{tabular}{cc}
      \resizebox{43mm}{!}{\includegraphics{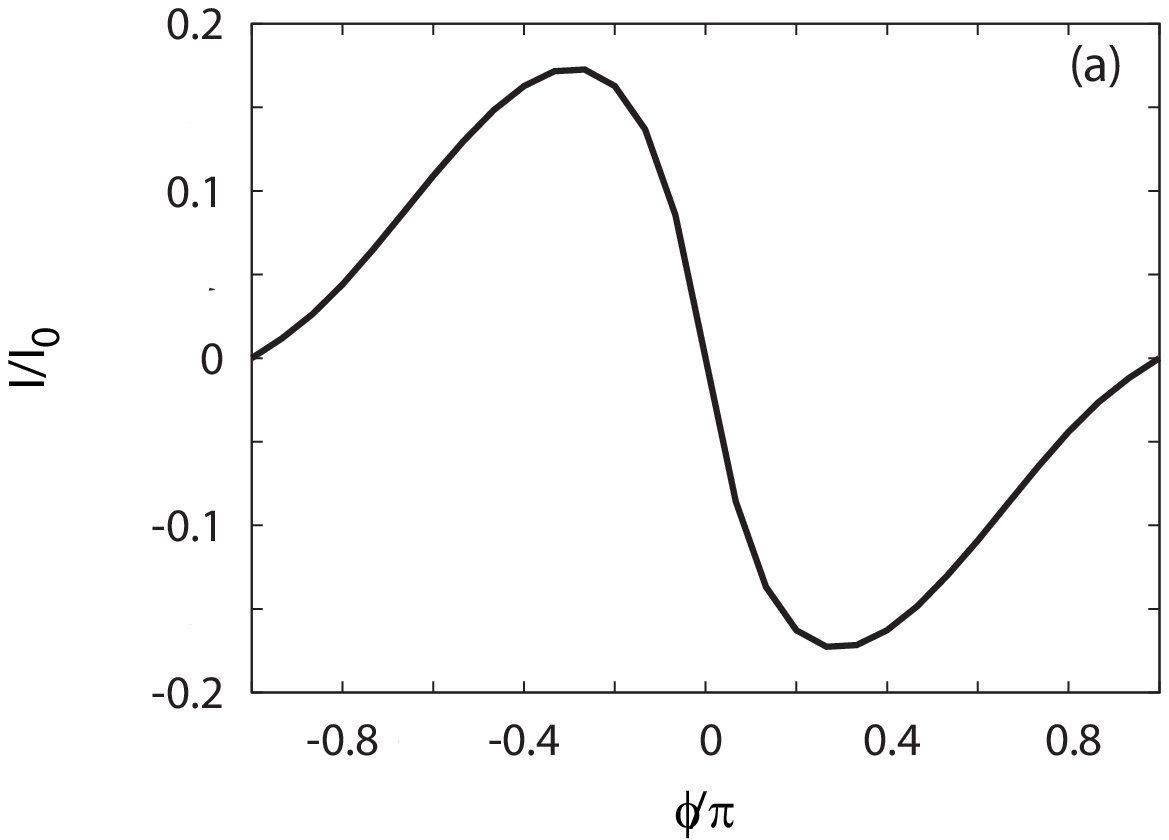}} &
      \resizebox{43mm}{!}{\includegraphics{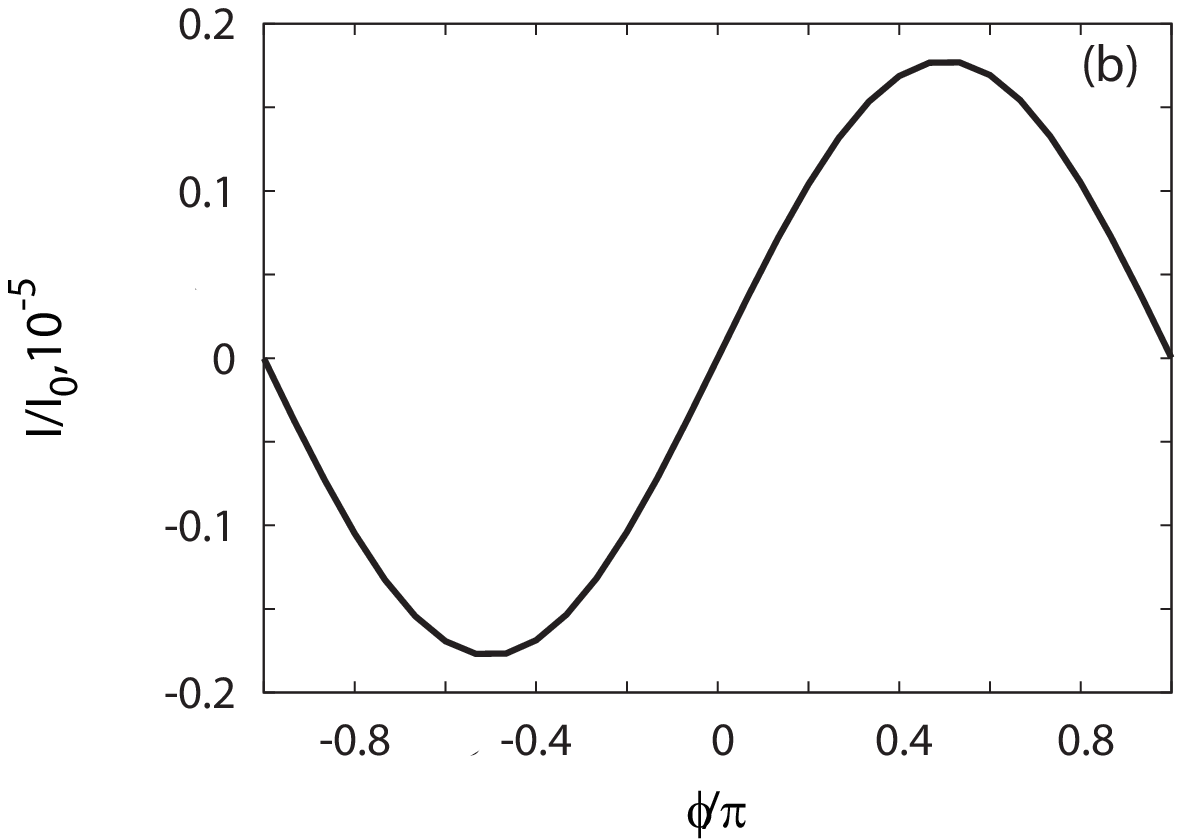}} \\
    \end{tabular}
\caption{The CPR of the $S/I/S_p$ Josephson junction with transport in z-direction for  a) the direct contact $I/S_p$; b) $N=3$ layers of an insulator atoms.}
    \label{R5}
  \end{center}
  \end{figure}

Let us summarize the obtained results of S/I/S$_{p}$ junction along (100) direction.
In the case of $s_{++}$ pairing symmetry in $S_{p}$, the junction is in $0$-junction only. On the other hand, in the case of the $s_{\pm}$ symmetry,
changing the interface hopping parameters, the size of the Fermi surface in $s$-wave superconductor and the insulating barrier length,
we can obtain $0$-, $\pi$- or $\phi$-junction. In the latter case, important feature of the S/I/S$_{p}$ junction is the existence of large second harmonic in CPR
in a broad parameter range, as illustrated in Fig. \ref{Harmonic}. The physical origin of large second harmonic is related to interband interference effects
in the $s_{\pm}$ pairing state. These effects manifest themselves in the formation of additional current-carrying surface bound states.

Next, we present the results of calculations of CPR in $S/I/S_p$ junction along $z$-axis using
the tight-binding Green's functions obtained in Sec. \ref{sec3}B.
We assume that the normal excitation spectrum in $S$ has the form  $\varepsilon_N=2t(\cos k_x + \cos k_y + \cos k_z) + \mu_N$ with hopping parameter $t=-0.3$ (eV) and chemical potential $\mu_N=0.6$ (eV).
For the chosen values of  hopping parameters and chemical potential the size of  Fermi surface in $S$ is sufficiently large and both electronic and hole pockets in a FeBS contribute to the Josephson current.
We choose hopping  $t_z=-0.1$ (eV) between the same orbitals on the nearest neighbor sites of FeBS along $z$-axis.
We assume that the $S/I$ interface is fully transparent and
the $I/S_p$ interface is characterized by the following set of hopping amplitudes: $\gamma_{1z}= \gamma_{2z}=0.17$.
As in the previous case, we consider the low temperature regime: $T/T^s_c \approx 0.02$.
In contrast to the case of (100) oriented  $S/I/S_p$ junction, only one of the FeBS bands contributes to  the Josephson current
at each fixed $k_{||}=(k_x,k_y)$ for transport in $z$-direction.

The CPR of $S/I/S_p$ junction along $z$-direction averaged over $k_{||}=(k_x,k_y)$ are plotted
in Fig.\ref{R5} for the direct contact (a) and for the case of $N=3$ insulating layers (b). In the direct contact,
the main contribution to the total Josephson current stems from electronic pockets.
This, the $S/I/S_p$ Josephson junction has ground state at $\pi$ phase difference (Fig.\ref{R5},a).
In the presence of the insulating barrier, the main contribution to the Josephson current stems from hole pockets
due to the suppression of the contributions from the regions with large $k_{||}$  to the total current.
As a result, the junction has ground state at zero phase difference (Fig.\ref{R5},b).

Modern microfabrication techniques make it possible to create dc SQUID loop with two different types of junctions, transparent and insulating one,
attached to a $c$-oriented FeBS.
Observation of $\pi$ phase shift in such device could provide crucial evidence for the $s_{\pm}$ symmetry in a FeBS.
Such experimental setup has been proposed  recently in Ref.\cite{Golubov2013}.
Important feature of $c$-oriented  $S/I/S_p$ Josephson junction is significant suppression
of the magnitude of the Josephson current in the case of long insulating layer (Fig.\ref{R5},b) compared to the direct contact (Fig.\ref{R5},a).
The suppression of the Josephson  current was observed
in recent Josephson tunneling experiments in a FeBS \cite{Siedel2011}.

 \begin{figure}
\begin{minipage}[h]{0.47\linewidth}
\center{\includegraphics[width=1\linewidth]{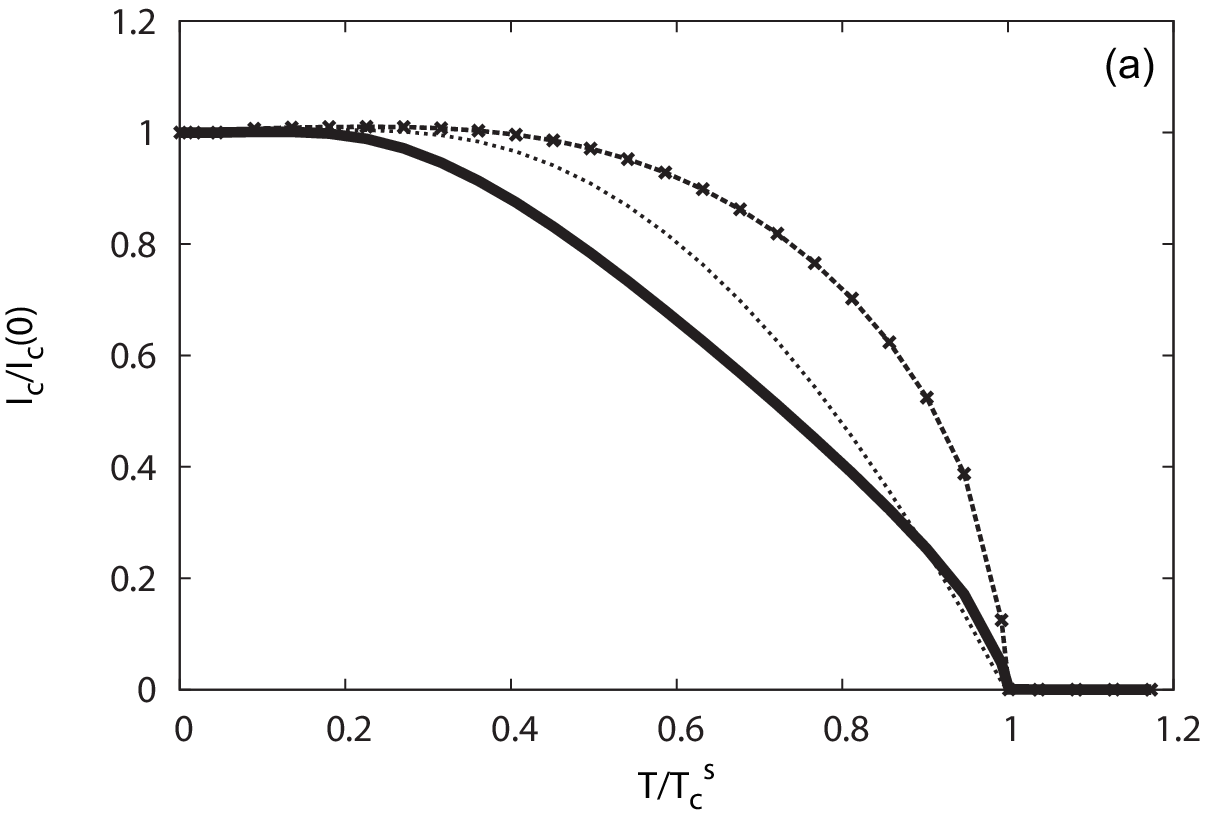}} \\
\end{minipage}
\hfill
\begin{minipage}[h]{0.47\linewidth}
\center{\includegraphics[width=1\linewidth]{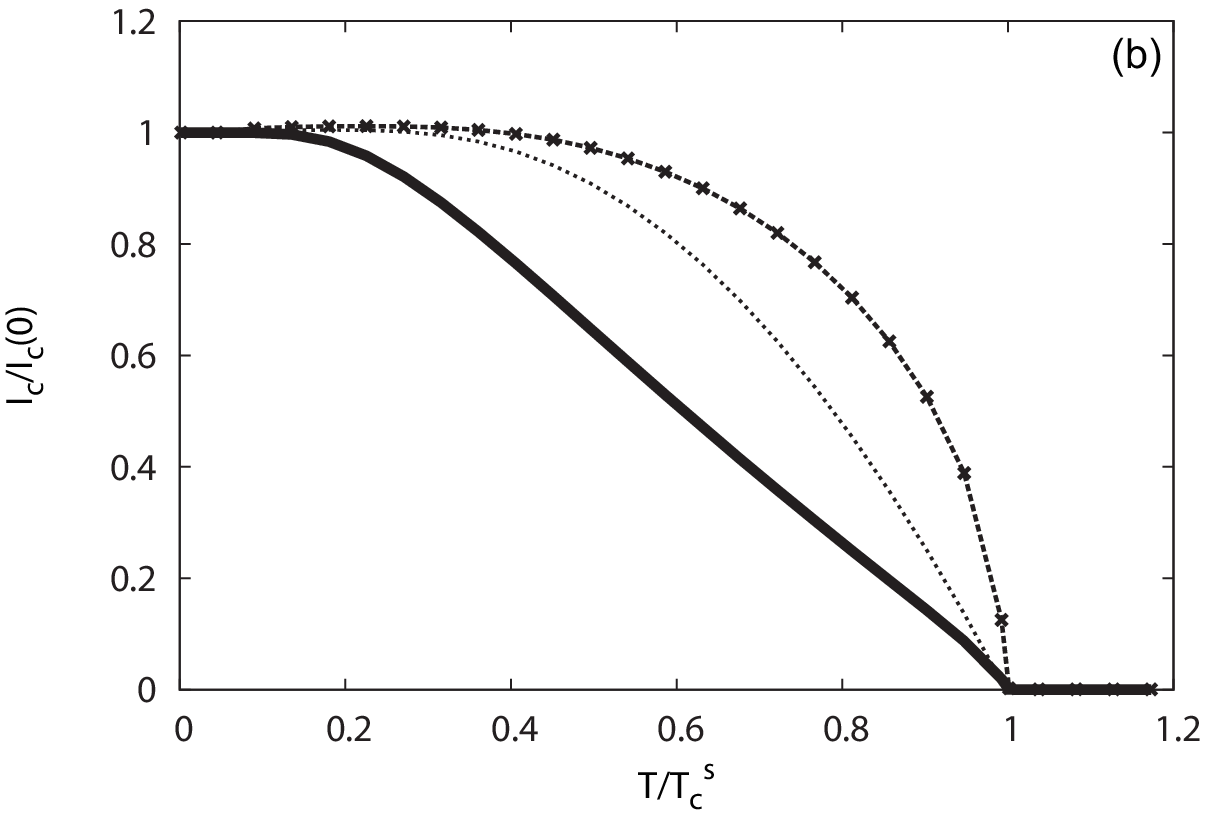}} \\
\end{minipage}
\vfill
\begin{minipage}[h]{0.47\linewidth}
\center{\includegraphics[width=1\linewidth]{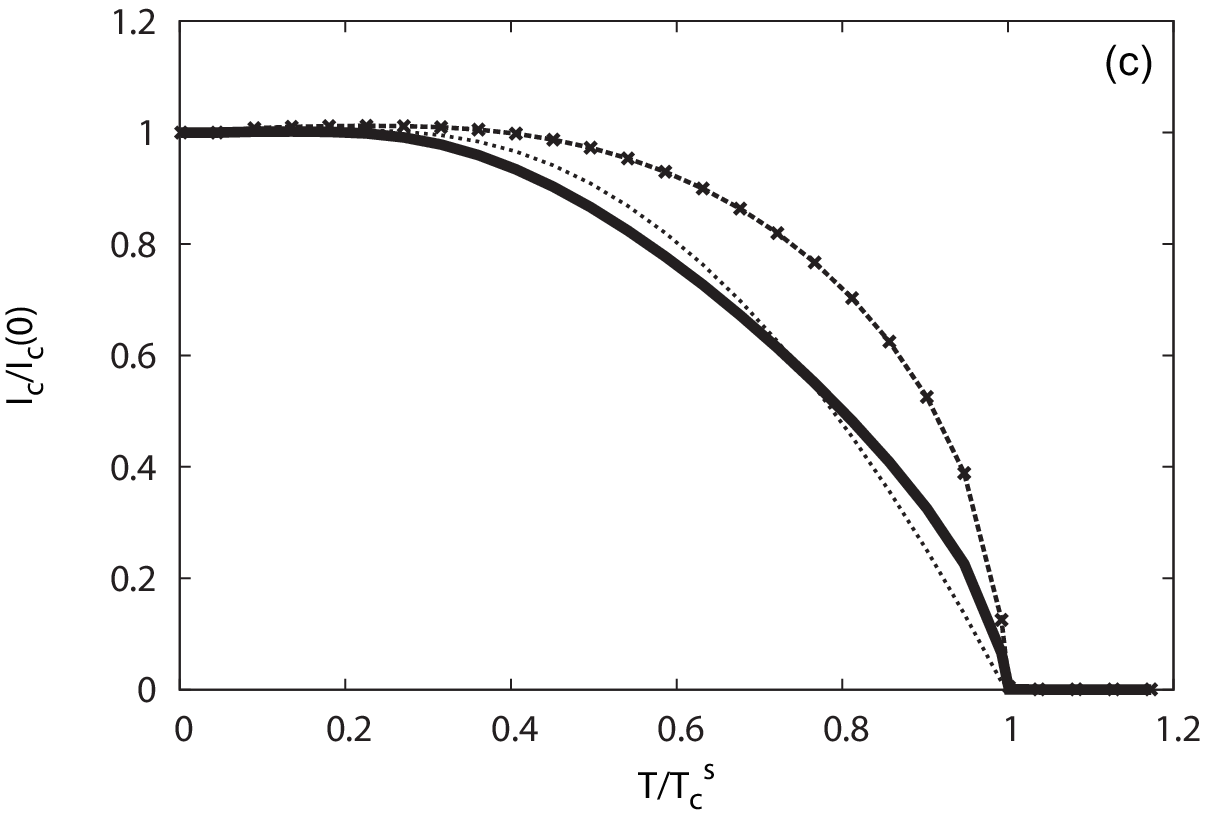}} \\
\end{minipage}
\hfill
\begin{minipage}[h]{0.47\linewidth}
\center{\includegraphics[width=1\linewidth]{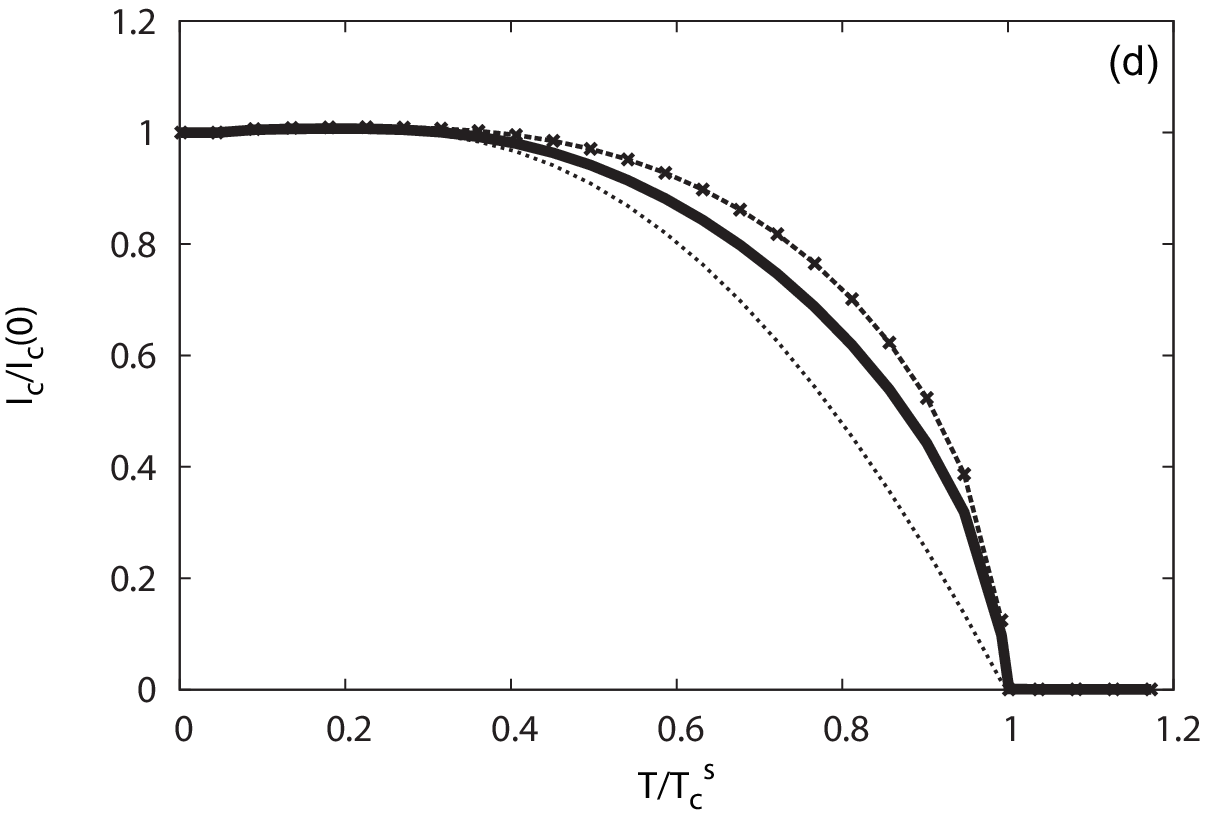}} \\
\end{minipage}
\caption{The temperature dependence of the maximum Josephson current of  the (100) oriented $S/I/S_p$ junction  for zero misorientation angle with respect to the interface. Solid line corresponds to the direct contact and line with crosses corresponds to  $N=3$  insulator layers. Dashed line corresponds to the Ambegaokar-Baratoff  temperature dependence of the maximum Josephson current in a conventional $S/I/S$ junction; (a) $\gamma_1=0.02, \gamma_2=0.2$; (b) $\gamma_1=0.02, \gamma_2=0.3$; (c) $\gamma_1=0.02, \gamma_2=0.4$; (d) $\gamma_1=0.2, \gamma_2=0.02$.}
\label{IcT}
\end{figure}

\subsection{Temperature dependencies of the Josephson critical current in $S/I/S_p$ junctions}

Temperature dependencies of the Josephson critical current in $S/I/S_p$ junctions were calculated
in the framework of the developed tight-binding Green's function approach.
To search for manifestations of unconventional pairing symmetry in FeBS, we considered the case of the $s_{\pm}$ pairing symmetry.
The results are shown in Fig.\ref{IcT} for the following choice of hopping parameters across the $I/S_p$ interface:
$\gamma_1=0.02,\gamma_2=0.2$ in Fig.\ref{IcT}(a),
$\gamma_1=0.02,\gamma_2=0.3$ in Fig.\ref{IcT}(b),
$\gamma_1=0.02,\gamma_2=0.4$ in Fig.\ref{IcT}(c)  and
$\gamma_1=0.2,\gamma_2=0.02$ in Fig.\ref{IcT}(d).

The solid lines in Fig.\ref{IcT} correspond to the the direct contact, the lines with crosses - to the $S/I/S_p$ junction with a thick insulating layer and the dotted lines
 show the Ambegaokar-Baratoff \cite{amb} temperature dependence for the Josephson critical current in a standard $S/I/S$ junction. One can see from Fig.\ref{IcT}(a)-(d) that the  Josephson current decreases with temperature more slowly in the case of the $S/I/S_p$ structure with long insulating layer compared to the $S/I/S$ junction in the whole considered parameter range. The behavior of the critical current in $S/I/S_p$ junctions with the direct contact depends on a choice of the hopping parameters at the $I/S_p$ interface. The most significant difference compared to the Ambegaokar-Baratoff temperature dependence occurs in the case of $\gamma_1=0.02,\gamma_2=0.3$ (Fig.\ref{IcT}(b)). This choice of hopping parameters corresponds to the realization of nontrivial phase dependence of the Josephson current with phase difference in the ground state at $\phi$ ($0<\phi<\pi$) (Fig.\ref{R2},a).

Our calculations demonstrate that the temperature dependencies of the Josephson critical current in z-axis $S/I/S_p$ junctions, both for the direct contact and for $N=3$  insulating layers, are quite close to each other. In both cases $I_c(T)$ falls down with temperature more slowly than in a standard $S/I/S$ tunnel junction.

\section{Experimental results}
\begin{figure}[h!]
\centerline{\includegraphics[width=8cm]{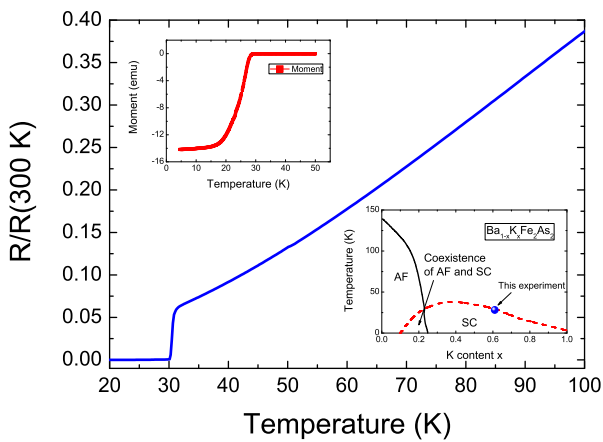}} 

\vspace{53mm}

\caption{Normalized resistance R/R(300K) of the Ba$_{0.4}$K$_{0.6}$(FeAs)$_2$ single crystals. Upper inset: zero-field cooled magnetization measurement. Lower inset: phase diagram of Ba$_{1-x}$K$_{x}$(FeAs)$_2$ \cite{Liu,Avci}. The blue symbol represents the samples studied in this work. T$_c$ has been determined from the magnetization curve shown in the upper inset.}\label{Res}
\end{figure}

{\label{Sec_Exp}}
The experiments were performed on Ba$_{0.4}$K$_{0.6}$(FeAs)$_2$ single crystals with T$_c \approx 30$ K. The samples were fabricated by the self-flux method. Firstly, precursor materials (BaAs, KAs and Fe$_2$As) were prepared by sintering elemental mixtures at $400^\circ$C$, 600^\circ$C and $700^\circ$C, respectively. After the careful weighing procedure, the starting precursors with a ratio of KAs:BaAs:Fe$_2$As =3.6:0.4:1 were loaded into an alumina crucible and then sealed in a tantalum tube under 1 atm of argon gas. By sealing the tube in an evacuated quartz tube, the chemicals were subsequently heated up to $1050^\circ$C and held for 5 hours. Then the furnace was cooled down to $900^\circ$C at a rate of $3^\circ$C/h and from $900^\circ$C to $600^\circ$C at $5^\circ$C/h. Finally the power of the furnace was shut off, and the samples were obtained by washing out the KAs flux. The EDS analysis showed that the effective composition was Ba$_{0.41}$K$_{0.61}$Fe$_{1.97}$As$_2$, very close to the nominal one. For this reason we will keep referring to the samples by using the nominal content. Figure \ref{Res} shows the normalized resistance, R/R(300K), from which it is possible to notice that R(40K)/R(300K)$\approx 0.09$, in very good agreement with Ref. \cite{Liu}. Moreover, since it has been shown that Ba$_{1-x}$K$_{x}$(FeAs)$_2$ compounds are clean over the whole doping range \cite{Liu}, we can exclude any significant effect of scattering on the measured Josephson current. The lower inset of Figure \ref{Res} reports the phase diagram for the K-doped Ba 122 materials \cite{Liu,Avci}. The point on the diagram representative of the samples experimentally investigated in this work, shown as a blue symbol, has been obtained from the magnetization curve reported in the upper inset of Figure \ref{Res} and matches very well with the corresponding one of the phase diagram. Finally, let us note that the samples studied here are far off the region of coexistence of antiferromagnetism and superconductivity. Hence, possible effects related to such a coexistence cannot play a role.

PbIn/Ba$_{1-x}$K$_{x}$(FeAs)$_2$ point-contact Josephson junctions were fabricated using Pb$_{0.7}$In$_{0.3}$ alloy (T$_c \approx 6.5$ K, as determined by the temperature at which the Josephson current vanishes) as the counterelectrode. A sharpened tip was used for injecting the current along the $c$-axis while a wedge-like one was employed for current injection along the $ab$-plane. The contacts were formed at low temperature by means of a differential micrometer.\\
Reproducible, non-hysteretic RSJ-like I-V characteristics were observed at low temperature. The junctions were then irradiated with microwaves by using a monopole antenna placed at the end of a semi-rigid coaxial cable. The occurrence of the Josephson effect was proved by the presence of microwave-induced current steps at voltages multiple of $\hbar \omega_{rf}/2e$, where $\omega_{rf}$ is the microwave frequency. Subsequently, the power dependence of the current steps was investigated.\\

\begin{figure}[h!]
\centerline{\includegraphics[width=8cm]{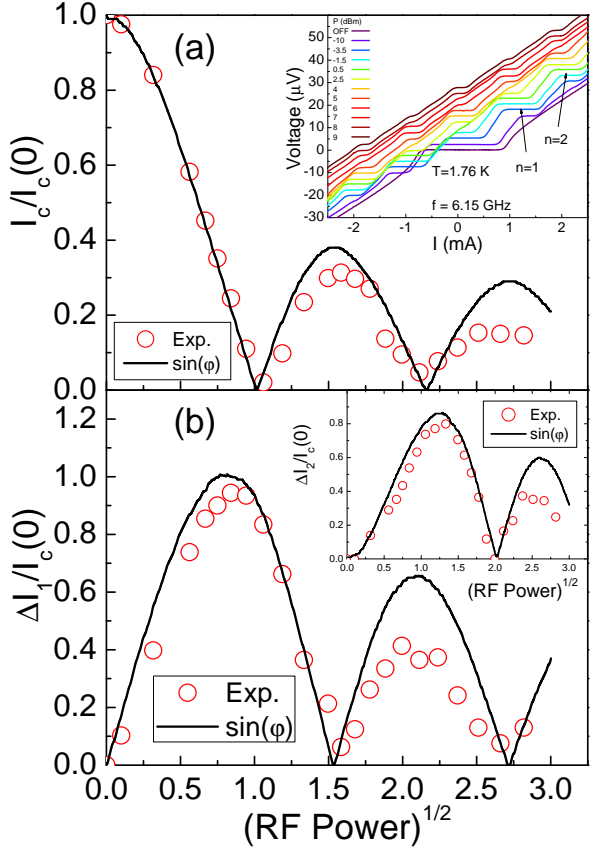}}
\caption{Pb$_{0.7}$In$_{0.3}$/Ba$_{0.4}$K$_{0.6}$(FeAs)$_2$ point-contact junctions with current injection along the $c$ axis. Panel (a), inset: subset of I-V curves at T=1.76 K irradiated with a 6.15 GHz rf frequency at different power levels. Main panel: normalized critical current as a function of the square root of the power. Panel (b): normalized amplitude of step 1 vs (RF Power)$^{1/2}$. Panel (b), inset: normalized amplitude of step 2 vs (RF Power)$^{1/2}$.}
\label{PC_c}
\end{figure}

Figure \ref{PC_c} shows the results obtained for a $c$-axis junction whose I$_c$R$_N$ product was about $12 \, \mu V$. The inset of panel (a) reports some of the I-V curves obtained at 1.76 K and in the presence of an rf irradiation of 6.15 GHz at different power levels. It can be seen that, as expected, the amplitude of both the critical current and of the higher-order steps modulates when changing the power. Panel (a) (symbols) shows the behavior of the critical current as a function of the square root of the power while panel (b) and the inset of panel (b) (symbols) report the amplitude of step 1 and 2, respectively. All the steps were normalized by the low-temperature critical current.\\
To describe the junction under microwave irradiation, the RSJ model is extended to the nonautonomous case with an rf current--source term \cite{barone}. For the results of Figure \ref{PC_c}, the model has been calculated supposing $I=I_c\sin(\varphi)$ as the current-phase relation and by using the parameter $\Omega=\hbar\omega_{rf}/2eI_cR_N=1$, as imposed by the experiment. Then, since the actual microwave power coupling with the junction is unknown, a scaling parameter for the power was used to fit the data, as it is usual in these cases \cite{Sellier}. Lines in panel (a), (b) and inset of panel (b) are the results of the calculations. It is worth noticing that the scaling parameter for the power is of course the same for all the current steps shown. It can be clearly seen that the agreement between the model and the experimental results is very good. This agrees also well with what shown in Figure \ref{R5} (a), where a dominant $\sin(\varphi)$ component has been predicted for the current-phase relation along the $z$ direction in a direct contact.\\

\begin{figure}[h!]
\centerline{\includegraphics[width=8cm]{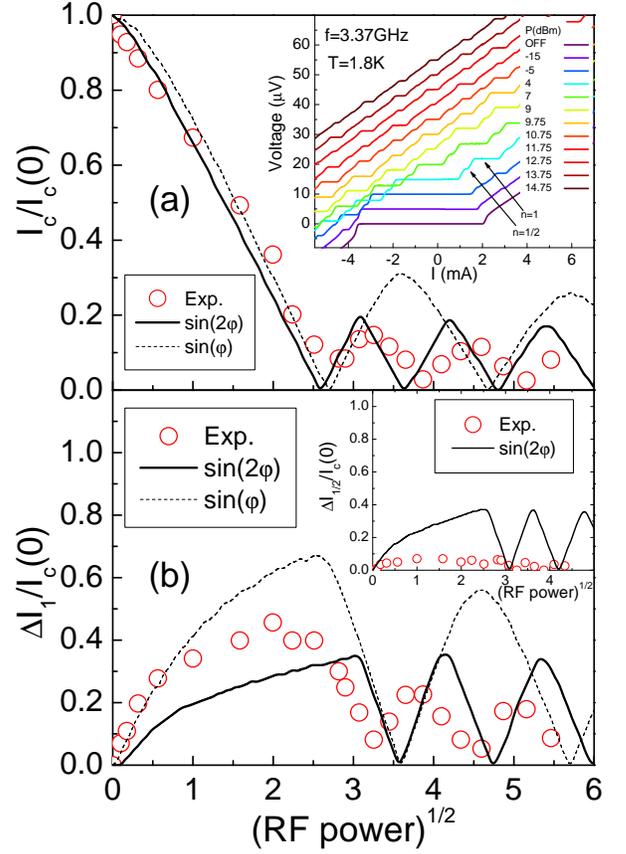}}
\caption{Pb$_{0.7}$In$_{0.3}$/Ba$_{0.4}$K$_{0.6}$(FeAs)$_2$ point-contact junctions with current injection along the $ab$ plane. Panel (a), inset: subset of I-V curves at T=1.8 K irradiated with a 3.37 GHz rf frequency at different power levels. Main panel: normalized critical current as a function of the square root of the power. Panel (b): normalized amplitude of step 1 vs (RF Power)$^{1/2}$. Panel (b), inset: normalized amplitude of step 1/2 vs (RF Power)$^{1/2}$.}
\label{PC_ab}
\end{figure}
Figure \ref{PC_ab} shows a typical result obtained for current injection along the $ab$-plane. The inset of panel (a) reports a subset of I-V curves measured at 1.8 K and under a microwave irradiation of 3.37 GHz at different power levels. The I$_c$R$_N$ product for the non-irradiated curve was approximately $15 \, \mu V$. The irradiated curves show the occurrence of current-induced steps at voltages $n\hbar \omega_{rf}/2e$, where $n$ is an integer, but also at $(n/2)\hbar \omega_{rf}/2e$, indicating the presence of a second-harmonic component in the current-phase relation. Also in this case the amplitude of the steps oscillates with increasing rf power.\\
Panel (a), panel (b) and inset of panel (b) report the behavior, as a function of the square root of the rf power, of the amplitude of the critical current, of step 1 and of step 1/2, respectively (symbols). The data were compared to the nonautonomous case with $\Omega = 0.42$, as determined by the experiment. The equation was first solved with $I=I_c\sin(\varphi)$. The result is shown in the figure as dashed lines. This solution clearly fails in reproducing the data in amplitude but especially in following the period of the steps oscillations. Besides, the fractional steps are of course not obtained. Therefore, a solution of the model with $I=I_c\sin(2\varphi)$ has been calculated as well and is shown as solid lines. In this case the fit, though not perfect, is quite close to the actual experimental behavior, especially for steps 0 and 1. Also in this case, for each expression of $I$, only one fitting parameter has been used for all the steps. The slight discrepancy between the model with a pure second-harmonic component and the experimental data suggests that the actual current-phase relation is not exactly $I=I_c\sin(2\varphi)$ but most probably a mixing of the first and second harmonic (see Figure \ref{Harmonic}). The presence of a further component in the current-phase relation can be inferred for example by the incomplete suppression of the first minimum of the supercurrent (panel (a)) and of the first step (panel (b)) \cite{Kleiner}, as well as by the larger amplitude of the theoretical step 1/2 in the inset of panel (b). Finally, it is worth recalling that there may be, in principle, other reasons for the appearance of subharmonic steps, but they can be excluded to play a role here \cite{belenov,Seidel,cuevas}. Possible accidental nodes in K-doped samples do not modify qualitatively CPR of the Josephson current because, as it was demonstrated in \cite{tanaka97}, CPR is modified qualitatively only in the case of sign-change of the order parameter. Possible nodes in K-doped samples do not imply sign-change of an order parameter.\\
These results indicate that a $\sin(2\varphi)$ component is highly dominant in the CPR of junctions with current injection along the $ab$-plane. As shown in Figure \ref{Harmonic}, this situation is predicted for a broad range of values of the hopping parameters in case of in-plane tunneling between a conventional superconductor and a multi-band superconductor with an $s\pm$-wave symmetry of the order parameter. Indeed, as reported in more detail in section \ref{sec4}, a large second-harmonic component in the CPR can occur as a consequence of interband interference effects within the $s\pm$-wave model.\\
Therefore, these experiments appear to be in good agreement with the theoretical calculations of the Josephson current presented here in case of an $s\pm$-wave symmetry of the order parameter. However, direct measurements of the current-phase relation are desirable, in order to catch finer details of the actual current-phase relation.\\

\section{Conclusion}

In this paper in the framework of tight-binding model we have proposed a microscopic theory describing Josephson tunneling in junctions with unusual multiband superconductors. Our theory takes into account not only the complex excitation spectrum of these superconductors, their multiband Fermi surface, interband and intervalley scattering at the boundaries, but also anisotropy and possible sign-changing of the order parameter in them. This theory has been applied to the calculation of the current-phase relation of the Josephson current and temperature dependence of the maximum Josephson current of a FeBS / spin-singlet $s$-wave single-orbital superconductor junction for
different orientation of the crystal axes of a FeBS by changing the length of an insulating layer. We have investigated experimentally PbIn/Ba$_{1-x}$K$_{x}$(FeAs)$_2$ point-contact Josephson junctions and based on our theory have demonstrated that $s_{\pm}$ scenario is more probable than $s_{++}$ in Ba$_{1-x}$K$_{x}$(FeAs)$_2$. A largely dominant second-harmonic component in the CPR has indeed been observed in case of current injection along the $ab$-plane, as predicted by the theory and shown in Figure \ref{Harmonic}. We have demonstrated theoretically that to measure the Josepshon current in the junction parallel to $c$-axis of a FeBS allows to distinguish the $s_{\pm}$-wave from $s_{++}$-wave in  FeBS. In the light of our theory, the recently proposed experimental set up  to determine the symmetry of the order parameter in a FeBS \cite{Golubov2013} has been confirmed to be plausible. It is interesting to note that our proposed theoretical scheme in the framework of tight-binding model technique can be used for calculations of the charge transport in structures with different unconventional and complex superconductors, such as other multiband superconductors\cite{Yada2014},
superconductor on the topological insulators \cite{fu08,tanaka09,Linder10},
and superconducting topological insulators \cite{sasaki11,yamakage12}.
Also it is interesting to focus on the properties of anomalous Green's function in terms of odd-frequency pairing\cite{Tanaka2007}
and its relevance to topological edge state
\cite{tanaka12},
since odd-frequency pairing and
Majorana fermion in the multi band system is a current topic now
\cite{Balatsky1,Balatsky2,fu08,Hao1,Hao2}.

\begin{acknowledgments}
We gratefully acknowledge T.M. Klapwijk and M.Y. Kupriyanov, D.J. Van Harlingen and P. Seidel for valuable discussions.
We thank W.K. Park, J. M. Atkinson Mora, R.W. Giannetta, and J. Ku for technical support.
This work was supported by the Russian Foundation for Basic
Research, projects N 13-02-01085, and N 14-02-31366-mol\_a, and 	
N 15-52-50054, the Ministry of Education and Science
of the Russian Federation  contract N 14.B25.31.0007 of 26 June 2013 and Grant No. 14.587.21.0006 (RFMEFI58714X0006), EU COST program MP1201,  the US National Science Foundation Division of Materials Research (DMR), Award 12-06766, and the U.S.-Italy Fulbright Commission for the Core Fulbright Visiting Scholar Program, during which the experimental results presented here were collected.
\end{acknowledgments}


\bibliography{biblio5}

\begin{thebibliography}{63}%
\makeatletter
\providecommand \@ifxundefined [1]{%
 \@ifx{#1\undefined}
}%
\providecommand \@ifnum [1]{%
 \ifnum #1\expandafter \@firstoftwo
 \else \expandafter \@secondoftwo
 \fi
}%
\providecommand \@ifx [1]{%
 \ifx #1\expandafter \@firstoftwo
 \else \expandafter \@secondoftwo
 \fi
}%
\providecommand \natexlab [1]{#1}%
\providecommand \enquote  [1]{``#1''}%
\providecommand \bibnamefont  [1]{#1}%
\providecommand \bibfnamefont [1]{#1}%
\providecommand \citenamefont [1]{#1}%
\providecommand \href@noop [0]{\@secondoftwo}%
\providecommand \href [0]{\begingroup \@sanitize@url \@href}%
\providecommand \@href[1]{\@@startlink{#1}\@@href}%
\providecommand \@@href[1]{\endgroup#1\@@endlink}%
\providecommand \@sanitize@url [0]{\catcode `\\12\catcode `\$12\catcode
  `\&12\catcode `\#12\catcode `\^12\catcode `\_12\catcode `\%12\relax}%
\providecommand \@@startlink[1]{}%
\providecommand \@@endlink[0]{}%
\providecommand \url  [0]{\begingroup\@sanitize@url \@url }%
\providecommand \@url [1]{\endgroup\@href {#1}{\urlprefix }}%
\providecommand \urlprefix  [0]{URL }%
\providecommand \Eprint [0]{\href }%
\providecommand \doibase [0]{http://dx.doi.org/}%
\providecommand \selectlanguage [0]{\@gobble}%
\providecommand \bibinfo  [0]{\@secondoftwo}%
\providecommand \bibfield  [0]{\@secondoftwo}%
\providecommand \translation [1]{[#1]}%
\providecommand \BibitemOpen [0]{}%
\providecommand \bibitemStop [0]{}%
\providecommand \bibitemNoStop [0]{.\EOS\space}%
\providecommand \EOS [0]{\spacefactor3000\relax}%
\providecommand \BibitemShut  [1]{\csname bibitem#1\endcsname}%
\let\auto@bib@innerbib\@empty
\bibitem [{\citenamefont {Wollman}\ \emph {et~al.}(1993)\citenamefont
  {Wollman}, \citenamefont {Van~Harlingen}, \citenamefont {Lee}, \citenamefont
  {Ginsberg},\ and\ \citenamefont {Leggett}}]{Wollman1993}%
  \BibitemOpen
  \bibfield  {author} {\bibinfo {author} {\bibfnamefont {D.~A.}\ \bibnamefont
  {Wollman}}, \bibinfo {author} {\bibfnamefont {D.~J.}\ \bibnamefont
  {Van~Harlingen}}, \bibinfo {author} {\bibfnamefont {W.~C.}\ \bibnamefont
  {Lee}}, \bibinfo {author} {\bibfnamefont {D.~M.}\ \bibnamefont {Ginsberg}}, \
  and\ \bibinfo {author} {\bibfnamefont {A.~J.}\ \bibnamefont {Leggett}},\
  }\href {\doibase 10.1103/PhysRevLett.71.2134} {\bibfield  {journal} {\bibinfo
   {journal} {Phys. Rev. Lett.}\ }\textbf {\bibinfo {volume} {71}},\ \bibinfo
  {pages} {2134} (\bibinfo {year} {1993})}\BibitemShut {NoStop}%
\bibitem [{\citenamefont {Tsuei}\ \emph {et~al.}(1994)\citenamefont {Tsuei},
  \citenamefont {Kirtley}, \citenamefont {Chi}, \citenamefont {Yu-Jahnes},
  \citenamefont {Gupta}, \citenamefont {Shaw}, \citenamefont {Sun},\ and\
  \citenamefont {Ketchen}}]{Tsuei1994}%
  \BibitemOpen
  \bibfield  {author} {\bibinfo {author} {\bibfnamefont {C.~C.}\ \bibnamefont
  {Tsuei}}, \bibinfo {author} {\bibfnamefont {J.~R.}\ \bibnamefont {Kirtley}},
  \bibinfo {author} {\bibfnamefont {C.~C.}\ \bibnamefont {Chi}}, \bibinfo
  {author} {\bibfnamefont {L.~S.}\ \bibnamefont {Yu-Jahnes}}, \bibinfo {author}
  {\bibfnamefont {A.}~\bibnamefont {Gupta}}, \bibinfo {author} {\bibfnamefont
  {T.}~\bibnamefont {Shaw}}, \bibinfo {author} {\bibfnamefont {J.~Z.}\
  \bibnamefont {Sun}}, \ and\ \bibinfo {author} {\bibfnamefont {M.~B.}\
  \bibnamefont {Ketchen}},\ }\href {\doibase 10.1103/PhysRevLett.73.593}
  {\bibfield  {journal} {\bibinfo  {journal} {Phys. Rev. Lett.}\ }\textbf
  {\bibinfo {volume} {73}},\ \bibinfo {pages} {593} (\bibinfo {year}
  {1994})}\BibitemShut {NoStop}%
\bibitem [{\citenamefont {Van~Harlingen}(1995)}]{VanHarlingen1995}%
  \BibitemOpen
  \bibfield  {author} {\bibinfo {author} {\bibfnamefont {D.~J.}\ \bibnamefont
  {Van~Harlingen}},\ }\href {\doibase 10.1103/RevModPhys.67.515} {\bibfield
  {journal} {\bibinfo  {journal} {Rev. Mod. Phys.}\ }\textbf {\bibinfo {volume}
  {67}},\ \bibinfo {pages} {515} (\bibinfo {year} {1995})}\BibitemShut
  {NoStop}%
\bibitem [{\citenamefont {Tsuei}\ and\ \citenamefont
  {Kirtley}(2000)}]{Tsuei2000}%
  \BibitemOpen
  \bibfield  {author} {\bibinfo {author} {\bibfnamefont {C.~C.}\ \bibnamefont
  {Tsuei}}\ and\ \bibinfo {author} {\bibfnamefont {J.~R.}\ \bibnamefont
  {Kirtley}},\ }\href {\doibase 10.1103/RevModPhys.72.969} {\bibfield
  {journal} {\bibinfo  {journal} {Rev. Mod. Phys.}\ }\textbf {\bibinfo {volume}
  {72}},\ \bibinfo {pages} {969} (\bibinfo {year} {2000})}\BibitemShut
  {NoStop}%
\bibitem [{\citenamefont {Tanaka}\ and\ \citenamefont
  {Kashiwaya}(1995)}]{tanaka1}%
  \BibitemOpen
  \bibfield  {author} {\bibinfo {author} {\bibfnamefont {Y.}~\bibnamefont
  {Tanaka}}\ and\ \bibinfo {author} {\bibfnamefont {S.}~\bibnamefont
  {Kashiwaya}},\ }\href {\doibase 10.1103/PhysRevLett.74.3451} {\bibfield
  {journal} {\bibinfo  {journal} {Phys. Rev. Lett.}\ }\textbf {\bibinfo
  {volume} {74}},\ \bibinfo {pages} {3451} (\bibinfo {year}
  {1995})}\BibitemShut {NoStop}%
\bibitem [{\citenamefont {Kashiwaya}\ and\ \citenamefont
  {Tanaka}(2000)}]{kashiwaya00}%
  \BibitemOpen
  \bibfield  {author} {\bibinfo {author} {\bibfnamefont {S.}~\bibnamefont
  {Kashiwaya}}\ and\ \bibinfo {author} {\bibfnamefont {Y.}~\bibnamefont
  {Tanaka}},\ }\href {http://stacks.iop.org/0034-4885/63/i=10/a=202} {\bibfield
   {journal} {\bibinfo  {journal} {Rep. Prog. Phys.}\ }\textbf {\bibinfo
  {volume} {63}},\ \bibinfo {pages} {1641} (\bibinfo {year}
  {2000})}\BibitemShut {NoStop}%
\bibitem [{\citenamefont {Hu}(1994)}]{Hu}%
  \BibitemOpen
  \bibfield  {author} {\bibinfo {author} {\bibfnamefont {C.-R.}\ \bibnamefont
  {Hu}},\ }\href {\doibase 10.1103/PhysRevLett.72.1526} {\bibfield  {journal}
  {\bibinfo  {journal} {Phys. Rev. Lett.}\ }\textbf {\bibinfo {volume} {72}},\
  \bibinfo {pages} {1526} (\bibinfo {year} {1994})}\BibitemShut {NoStop}%
\bibitem [{\citenamefont {Barash}\ \emph {et~al.}(1996)\citenamefont {Barash},
  \citenamefont {Burkhardt},\ and\ \citenamefont {Rainer}}]{Barash}%
  \BibitemOpen
  \bibfield  {author} {\bibinfo {author} {\bibfnamefont {Y.~S.}\ \bibnamefont
  {Barash}}, \bibinfo {author} {\bibfnamefont {H.}~\bibnamefont {Burkhardt}}, \
  and\ \bibinfo {author} {\bibfnamefont {D.}~\bibnamefont {Rainer}},\ }\href
  {\doibase 10.1103/PhysRevLett.77.4070} {\bibfield  {journal} {\bibinfo
  {journal} {Phys. Rev. Lett.}\ }\textbf {\bibinfo {volume} {77}},\ \bibinfo
  {pages} {4070} (\bibinfo {year} {1996})}\BibitemShut {NoStop}%
\bibitem [{\citenamefont {Tanaka}\ and\ \citenamefont
  {Kashiwaya}(1997)}]{tanaka97}%
  \BibitemOpen
  \bibfield  {author} {\bibinfo {author} {\bibfnamefont {Y.}~\bibnamefont
  {Tanaka}}\ and\ \bibinfo {author} {\bibfnamefont {S.}~\bibnamefont
  {Kashiwaya}},\ }\href {\doibase 10.1103/PhysRevB.56.892} {\bibfield
  {journal} {\bibinfo  {journal} {Phys. Rev. B}\ }\textbf {\bibinfo {volume}
  {56}},\ \bibinfo {pages} {892} (\bibinfo {year} {1997})}\BibitemShut
  {NoStop}%
\bibitem [{\citenamefont {Maeno}\ \emph {et~al.}(1994)\citenamefont {Maeno},
  \citenamefont {Hashimoto}, \citenamefont {Yoshida}, \citenamefont
  {Nishizaki}, \citenamefont {Fujita}, \citenamefont {Bednorz},\ and\
  \citenamefont {Lichtenberg}}]{Maeno1994}%
  \BibitemOpen
  \bibfield  {author} {\bibinfo {author} {\bibfnamefont {Y.}~\bibnamefont
  {Maeno}}, \bibinfo {author} {\bibfnamefont {H.}~\bibnamefont {Hashimoto}},
  \bibinfo {author} {\bibfnamefont {K.}~\bibnamefont {Yoshida}}, \bibinfo
  {author} {\bibfnamefont {S.}~\bibnamefont {Nishizaki}}, \bibinfo {author}
  {\bibfnamefont {T.}~\bibnamefont {Fujita}}, \bibinfo {author} {\bibfnamefont
  {J.~G.}\ \bibnamefont {Bednorz}}, \ and\ \bibinfo {author} {\bibfnamefont
  {F.}~\bibnamefont {Lichtenberg}},\ }\href@noop {} {\bibfield  {journal}
  {\bibinfo  {journal} {Nature}\ }\textbf {\bibinfo {volume} {372}},\ \bibinfo
  {pages} {532} (\bibinfo {year} {1994})}\BibitemShut {NoStop}%
\bibitem [{\citenamefont {Mao}\ \emph {et~al.}(2001)\citenamefont {Mao},
  \citenamefont {Nelson}, \citenamefont {Jin}, \citenamefont {Liu},\ and\
  \citenamefont {Maeno}}]{Mao2001}%
  \BibitemOpen
  \bibfield  {author} {\bibinfo {author} {\bibfnamefont {Z.~Q.}\ \bibnamefont
  {Mao}}, \bibinfo {author} {\bibfnamefont {K.~D.}\ \bibnamefont {Nelson}},
  \bibinfo {author} {\bibfnamefont {R.}~\bibnamefont {Jin}}, \bibinfo {author}
  {\bibfnamefont {Y.}~\bibnamefont {Liu}}, \ and\ \bibinfo {author}
  {\bibfnamefont {Y.}~\bibnamefont {Maeno}},\ }\href@noop {} {\bibfield
  {journal} {\bibinfo  {journal} {Phys. Rev. Lett.}\ }\textbf {\bibinfo
  {volume} {87}},\ \bibinfo {pages} {037003} (\bibinfo {year}
  {2001})}\BibitemShut {NoStop}%
\bibitem [{\citenamefont {Kashiwaya}\ \emph {et~al.}(2011)\citenamefont
  {Kashiwaya}, \citenamefont {Kashiwaya}, \citenamefont {Kambara},
  \citenamefont {Furuta}, \citenamefont {Yaguchi}, \citenamefont {Tanaka},\
  and\ \citenamefont {Maeno}}]{Kashiwaya11}%
  \BibitemOpen
  \bibfield  {author} {\bibinfo {author} {\bibfnamefont {S.}~\bibnamefont
  {Kashiwaya}}, \bibinfo {author} {\bibfnamefont {H.}~\bibnamefont
  {Kashiwaya}}, \bibinfo {author} {\bibfnamefont {H.}~\bibnamefont {Kambara}},
  \bibinfo {author} {\bibfnamefont {T.}~\bibnamefont {Furuta}}, \bibinfo
  {author} {\bibfnamefont {H.}~\bibnamefont {Yaguchi}}, \bibinfo {author}
  {\bibfnamefont {Y.}~\bibnamefont {Tanaka}}, \ and\ \bibinfo {author}
  {\bibfnamefont {Y.}~\bibnamefont {Maeno}},\ }\href@noop {} {\bibfield
  {journal} {\bibinfo  {journal} {Phys. Rev. Lett.}\ }\textbf {\bibinfo
  {volume} {107}},\ \bibinfo {pages} {077003} (\bibinfo {year}
  {2011})}\BibitemShut {NoStop}%
\bibitem [{\citenamefont {Kamihara}\ \emph {et~al.}(2008)\citenamefont
  {Kamihara}, \citenamefont {Watanabe}, \citenamefont {Hirano},\ and\
  \citenamefont {Hosono}}]{kam}%
  \BibitemOpen
  \bibfield  {author} {\bibinfo {author} {\bibfnamefont {Y.}~\bibnamefont
  {Kamihara}}, \bibinfo {author} {\bibfnamefont {T.}~\bibnamefont {Watanabe}},
  \bibinfo {author} {\bibfnamefont {M.}~\bibnamefont {Hirano}}, \ and\ \bibinfo
  {author} {\bibfnamefont {H.}~\bibnamefont {Hosono}},\ }\href {\doibase
  10.1021/ja800073m} {\bibfield  {journal} {\bibinfo  {journal} {Journal of the
  American Chemical Society}\ }\textbf {\bibinfo {volume} {130}},\ \bibinfo
  {pages} {3296} (\bibinfo {year} {2008})},\ \Eprint
  {http://arxiv.org/abs/http://pubs.acs.org/doi/pdf/10.1021/ja800073m}
  {http://pubs.acs.org/doi/pdf/10.1021/ja800073m} \BibitemShut {NoStop}%
\bibitem [{\citenamefont {Hor}\ \emph {et~al.}(2010)\citenamefont {Hor},
  \citenamefont {Williams}, \citenamefont {Checkelsky}, \citenamefont
  {Roushan}, \citenamefont {Seo}, \citenamefont {Xu}, \citenamefont
  {Zandbergen}, \citenamefont {Yazdani}, \citenamefont {Ong},\ and\
  \citenamefont {Cava}}]{hor10}%
  \BibitemOpen
  \bibfield  {author} {\bibinfo {author} {\bibfnamefont {Y.~S.}\ \bibnamefont
  {Hor}}, \bibinfo {author} {\bibfnamefont {A.~J.}\ \bibnamefont {Williams}},
  \bibinfo {author} {\bibfnamefont {J.~G.}\ \bibnamefont {Checkelsky}},
  \bibinfo {author} {\bibfnamefont {P.}~\bibnamefont {Roushan}}, \bibinfo
  {author} {\bibfnamefont {J.}~\bibnamefont {Seo}}, \bibinfo {author}
  {\bibfnamefont {Q.}~\bibnamefont {Xu}}, \bibinfo {author} {\bibfnamefont
  {H.~W.}\ \bibnamefont {Zandbergen}}, \bibinfo {author} {\bibfnamefont
  {A.}~\bibnamefont {Yazdani}}, \bibinfo {author} {\bibfnamefont {N.~P.}\
  \bibnamefont {Ong}}, \ and\ \bibinfo {author} {\bibfnamefont {R.~J.}\
  \bibnamefont {Cava}},\ }\href {\doibase 10.1103/PhysRevLett.104.057001}
  {\bibfield  {journal} {\bibinfo  {journal} {Phys. Rev. Lett.}\ }\textbf
  {\bibinfo {volume} {104}},\ \bibinfo {pages} {057001} (\bibinfo {year}
  {2010})}\BibitemShut {NoStop}%
\bibitem [{\citenamefont {Sasaki}\ \emph {et~al.}(2011)\citenamefont {Sasaki},
  \citenamefont {Kriener}, \citenamefont {Segawa}, \citenamefont {Yada},
  \citenamefont {Tanaka}, \citenamefont {Sato},\ and\ \citenamefont
  {Ando}}]{sasaki11}%
  \BibitemOpen
  \bibfield  {author} {\bibinfo {author} {\bibfnamefont {S.}~\bibnamefont
  {Sasaki}}, \bibinfo {author} {\bibfnamefont {M.}~\bibnamefont {Kriener}},
  \bibinfo {author} {\bibfnamefont {K.}~\bibnamefont {Segawa}}, \bibinfo
  {author} {\bibfnamefont {K.}~\bibnamefont {Yada}}, \bibinfo {author}
  {\bibfnamefont {Y.}~\bibnamefont {Tanaka}}, \bibinfo {author} {\bibfnamefont
  {M.}~\bibnamefont {Sato}}, \ and\ \bibinfo {author} {\bibfnamefont
  {Y.}~\bibnamefont {Ando}},\ }\href {\doibase 10.1103/PhysRevLett.107.217001}
  {\bibfield  {journal} {\bibinfo  {journal} {Phys. Rev. Lett.}\ }\textbf
  {\bibinfo {volume} {107}},\ \bibinfo {pages} {217001} (\bibinfo {year}
  {2011})}\BibitemShut {NoStop}%
\bibitem [{\citenamefont {Ara\'ujo}\ and\ \citenamefont
  {Sacramento}(2009)}]{sacr}%
  \BibitemOpen
  \bibfield  {author} {\bibinfo {author} {\bibfnamefont {M.~A.~N.}\
  \bibnamefont {Ara\'ujo}}\ and\ \bibinfo {author} {\bibfnamefont {P.~D.}\
  \bibnamefont {Sacramento}},\ }\href {\doibase 10.1103/PhysRevB.79.174529}
  {\bibfield  {journal} {\bibinfo  {journal} {Phys. Rev. B}\ }\textbf {\bibinfo
  {volume} {79}},\ \bibinfo {pages} {174529} (\bibinfo {year}
  {2009})}\BibitemShut {NoStop}%
\bibitem [{\citenamefont {Burmistrova}\ and\ \citenamefont
  {Devyatov}(2012{\natexlab{a}})}]{dev1}%
  \BibitemOpen
  \bibfield  {author} {\bibinfo {author} {\bibfnamefont {A.~V.}\ \bibnamefont
  {Burmistrova}}\ and\ \bibinfo {author} {\bibfnamefont {I.~A.}\ \bibnamefont
  {Devyatov}},\ }\href@noop {} {\bibfield  {journal} {\bibinfo  {journal} {JETP
  Letters}\ }\textbf {\bibinfo {volume} {95}},\ \bibinfo {pages} {263}
  (\bibinfo {year} {2012}{\natexlab{a}})}\BibitemShut {NoStop}%
\bibitem [{\citenamefont {Sperstad}\ \emph {et~al.}(2009)\citenamefont
  {Sperstad}, \citenamefont {Linder},\ and\ \citenamefont {Sudb\o{}}}]{lind}%
  \BibitemOpen
  \bibfield  {author} {\bibinfo {author} {\bibfnamefont {I.~B.}\ \bibnamefont
  {Sperstad}}, \bibinfo {author} {\bibfnamefont {J.}~\bibnamefont {Linder}}, \
  and\ \bibinfo {author} {\bibfnamefont {A.}~\bibnamefont {Sudb\o{}}},\ }\href
  {\doibase 10.1103/PhysRevB.80.144507} {\bibfield  {journal} {\bibinfo
  {journal} {Phys. Rev. B}\ }\textbf {\bibinfo {volume} {80}},\ \bibinfo
  {pages} {144507} (\bibinfo {year} {2009})}\BibitemShut {NoStop}%
\bibitem [{\citenamefont {Golubov}\ \emph {et~al.}(2009)\citenamefont
  {Golubov}, \citenamefont {Brinkman}, \citenamefont {Tanaka}, \citenamefont
  {Mazin},\ and\ \citenamefont {Dolgov}}]{gol}%
  \BibitemOpen
  \bibfield  {author} {\bibinfo {author} {\bibfnamefont {A.~A.}\ \bibnamefont
  {Golubov}}, \bibinfo {author} {\bibfnamefont {A.}~\bibnamefont {Brinkman}},
  \bibinfo {author} {\bibfnamefont {Y.}~\bibnamefont {Tanaka}}, \bibinfo
  {author} {\bibfnamefont {I.~I.}\ \bibnamefont {Mazin}}, \ and\ \bibinfo
  {author} {\bibfnamefont {O.~V.}\ \bibnamefont {Dolgov}},\ }\href {\doibase
  10.1103/PhysRevLett.103.077003} {\bibfield  {journal} {\bibinfo  {journal}
  {Phys. Rev. Lett.}\ }\textbf {\bibinfo {volume} {103}},\ \bibinfo {pages}
  {077003} (\bibinfo {year} {2009})}\BibitemShut {NoStop}%
\bibitem [{\citenamefont {Devyatov}\ \emph {et~al.}(2010)\citenamefont
  {Devyatov}, \citenamefont {Romashka},\ and\ \citenamefont
  {Burmistrova}}]{rom}%
  \BibitemOpen
  \bibfield  {author} {\bibinfo {author} {\bibfnamefont {I.~A.}\ \bibnamefont
  {Devyatov}}, \bibinfo {author} {\bibfnamefont {M.~Y.}\ \bibnamefont
  {Romashka}}, \ and\ \bibinfo {author} {\bibfnamefont {A.~V.}\ \bibnamefont
  {Burmistrova}},\ }\href@noop {} {\bibfield  {journal} {\bibinfo  {journal}
  {JETP Letters}\ }\textbf {\bibinfo {volume} {91}},\ \bibinfo {pages} {297}
  (\bibinfo {year} {2010})}\BibitemShut {NoStop}%
\bibitem [{\citenamefont {Burmistrova}\ \emph
  {et~al.}(2011{\natexlab{a}})\citenamefont {Burmistrova}, \citenamefont
  {Karminskaya},\ and\ \citenamefont {Devyatov}}]{kar1}%
  \BibitemOpen
  \bibfield  {author} {\bibinfo {author} {\bibfnamefont {A.~V.}\ \bibnamefont
  {Burmistrova}}, \bibinfo {author} {\bibfnamefont {T.~Y.}\ \bibnamefont
  {Karminskaya}}, \ and\ \bibinfo {author} {\bibfnamefont {I.~A.}\ \bibnamefont
  {Devyatov}},\ }\href@noop {} {\bibfield  {journal} {\bibinfo  {journal} {JETP
  Letters}\ }\textbf {\bibinfo {volume} {93}},\ \bibinfo {pages} {133}
  (\bibinfo {year} {2011}{\natexlab{a}})}\BibitemShut {NoStop}%
\bibitem [{\citenamefont {Burmistrova}\ \emph
  {et~al.}(2011{\natexlab{b}})\citenamefont {Burmistrova}, \citenamefont
  {Devyatov}, \citenamefont {Kupriyanov},\ and\ \citenamefont
  {Karminskaya}}]{kar2}%
  \BibitemOpen
  \bibfield  {author} {\bibinfo {author} {\bibfnamefont {A.~V.}\ \bibnamefont
  {Burmistrova}}, \bibinfo {author} {\bibfnamefont {I.~A.}\ \bibnamefont
  {Devyatov}}, \bibinfo {author} {\bibfnamefont {M.~Y.}\ \bibnamefont
  {Kupriyanov}}, \ and\ \bibinfo {author} {\bibfnamefont {T.~Y.}\ \bibnamefont
  {Karminskaya}},\ }\href@noop {} {\bibfield  {journal} {\bibinfo  {journal}
  {JETP Letters}\ }\textbf {\bibinfo {volume} {93}},\ \bibinfo {pages} {203}
  (\bibinfo {year} {2011}{\natexlab{b}})}\BibitemShut {NoStop}%
\bibitem [{\citenamefont {Burmistrova}\ and\ \citenamefont
  {Devyatov}(2012{\natexlab{b}})}]{bcrus}%
  \BibitemOpen
  \bibfield  {author} {\bibinfo {author} {\bibfnamefont {A.~V.}\ \bibnamefont
  {Burmistrova}}\ and\ \bibinfo {author} {\bibfnamefont {I.~A.}\ \bibnamefont
  {Devyatov}},\ }\href@noop {} {\bibfield  {journal} {\bibinfo  {journal} {JETP
  Letters}\ }\textbf {\bibinfo {volume} {96}},\ \bibinfo {pages} {391}
  (\bibinfo {year} {2012}{\natexlab{b}})}\BibitemShut {NoStop}%
\bibitem [{\citenamefont {Burmistrova}\ \emph {et~al.}(2013)\citenamefont
  {Burmistrova}, \citenamefont {Devyatov}, \citenamefont {Golubov},
  \citenamefont {Yada},\ and\ \citenamefont {Tanaka}}]{Burmistrova2013}%
  \BibitemOpen
  \bibfield  {author} {\bibinfo {author} {\bibfnamefont {A.~V.}\ \bibnamefont
  {Burmistrova}}, \bibinfo {author} {\bibfnamefont {I.~A.}\ \bibnamefont
  {Devyatov}}, \bibinfo {author} {\bibfnamefont {A.~A.}\ \bibnamefont
  {Golubov}}, \bibinfo {author} {\bibfnamefont {K.}~\bibnamefont {Yada}}, \
  and\ \bibinfo {author} {\bibfnamefont {Y.}~\bibnamefont {Tanaka}},\
  }\href@noop {} {\bibfield  {journal} {\bibinfo  {journal} {Journal of the
  Physical Society of Japan}\ }\textbf {\bibinfo {volume} {82}},\ \bibinfo
  {pages} {034716} (\bibinfo {year} {2013})}\BibitemShut {NoStop}%
\bibitem [{\citenamefont {Berg}\ \emph {et~al.}(2011)\citenamefont {Berg},
  \citenamefont {Lindner},\ and\ \citenamefont {Pereg-Barnea}}]{ber}%
  \BibitemOpen
  \bibfield  {author} {\bibinfo {author} {\bibfnamefont {E.}~\bibnamefont
  {Berg}}, \bibinfo {author} {\bibfnamefont {N.~H.}\ \bibnamefont {Lindner}}, \
  and\ \bibinfo {author} {\bibfnamefont {T.}~\bibnamefont {Pereg-Barnea}},\
  }\href {\doibase 10.1103/PhysRevLett.106.147003} {\bibfield  {journal}
  {\bibinfo  {journal} {Phys. Rev. Lett.}\ }\textbf {\bibinfo {volume} {106}},\
  \bibinfo {pages} {147003} (\bibinfo {year} {2011})}\BibitemShut {NoStop}%
\bibitem [{\citenamefont {Chen}\ \emph {et~al.}(2009)\citenamefont {Chen},
  \citenamefont {Ma}, \citenamefont {Lu},\ and\ \citenamefont {Zhang}}]{chen}%
  \BibitemOpen
  \bibfield  {author} {\bibinfo {author} {\bibfnamefont {W.-Q.}\ \bibnamefont
  {Chen}}, \bibinfo {author} {\bibfnamefont {F.}~\bibnamefont {Ma}}, \bibinfo
  {author} {\bibfnamefont {Z.-Y.}\ \bibnamefont {Lu}}, \ and\ \bibinfo {author}
  {\bibfnamefont {F.-C.}\ \bibnamefont {Zhang}},\ }\href {\doibase
  10.1103/PhysRevLett.103.207001} {\bibfield  {journal} {\bibinfo  {journal}
  {Phys. Rev. Lett.}\ }\textbf {\bibinfo {volume} {103}},\ \bibinfo {pages}
  {207001} (\bibinfo {year} {2009})}\BibitemShut {NoStop}%
\bibitem [{\citenamefont {Koshelev}(2012)}]{koshelev2012}%
  \BibitemOpen
  \bibfield  {author} {\bibinfo {author} {\bibfnamefont {A.~E.}\ \bibnamefont
  {Koshelev}},\ }\href {\doibase 10.1103/PhysRevB.86.214502} {\bibfield
  {journal} {\bibinfo  {journal} {Phys. Rev. B}\ }\textbf {\bibinfo {volume}
  {86}},\ \bibinfo {pages} {214502} (\bibinfo {year} {2012})}\BibitemShut
  {NoStop}%
\bibitem [{\citenamefont {Golubov}\ and\ \citenamefont
  {Mazin}(2013)}]{Golubov2013}%
  \BibitemOpen
  \bibfield  {author} {\bibinfo {author} {\bibfnamefont {A.~A.}\ \bibnamefont
  {Golubov}}\ and\ \bibinfo {author} {\bibfnamefont {I.~I.}\ \bibnamefont
  {Mazin}},\ }\href@noop {} {\bibfield  {journal} {\bibinfo  {journal} {Appl.
  Phys. Lett.}\ }\textbf {\bibinfo {volume} {102}},\ \bibinfo {pages} {032601}
  (\bibinfo {year} {2013})}\BibitemShut {NoStop}%
\bibitem [{\citenamefont {Burmistrova}\ and\ \citenamefont
  {Devyatov}(2014)}]{EPL2014}%
  \BibitemOpen
  \bibfield  {author} {\bibinfo {author} {\bibfnamefont {A.~V.}\ \bibnamefont
  {Burmistrova}}\ and\ \bibinfo {author} {\bibfnamefont {I.~A.}\ \bibnamefont
  {Devyatov}},\ }\href {http://stacks.iop.org/0295-5075/107/i=6/a=67006}
  {\bibfield  {journal} {\bibinfo  {journal} {EPL (Europhysics Letters)}\
  }\textbf {\bibinfo {volume} {107}},\ \bibinfo {pages} {67006} (\bibinfo
  {year} {2014})}\BibitemShut {NoStop}%
\bibitem [{\citenamefont {Beenakker}\ and\ \citenamefont {van
  Houten}(1991)}]{Beenakker1991}%
  \BibitemOpen
  \bibfield  {author} {\bibinfo {author} {\bibfnamefont {C.~W.~J.}\
  \bibnamefont {Beenakker}}\ and\ \bibinfo {author} {\bibfnamefont
  {H.}~\bibnamefont {van Houten}},\ }\href {\doibase
  10.1103/PhysRevLett.66.3056} {\bibfield  {journal} {\bibinfo  {journal}
  {Phys. Rev. Lett.}\ }\textbf {\bibinfo {volume} {66}},\ \bibinfo {pages}
  {3056} (\bibinfo {year} {1991})}\BibitemShut {NoStop}%
\bibitem [{\citenamefont {Kulik}\ and\ \citenamefont
  {Omel'yanchuk}(1977)}]{Kulik1977}%
  \BibitemOpen
  \bibfield  {author} {\bibinfo {author} {\bibfnamefont {I.}~\bibnamefont
  {Kulik}}\ and\ \bibinfo {author} {\bibfnamefont {A.~N.}\ \bibnamefont
  {Omel'yanchuk}},\ }\href@noop {} {\bibfield  {journal} {\bibinfo  {journal}
  {Soviet Journal of Low Temperature Physics}\ }\textbf {\bibinfo {volume}
  {3}},\ \bibinfo {pages} {459} (\bibinfo {year} {1977})}\BibitemShut {NoStop}%
\bibitem [{\citenamefont {Kulik}\ and\ \citenamefont
  {Omel'yanchuk}(1978)}]{Kulik1978}%
  \BibitemOpen
  \bibfield  {author} {\bibinfo {author} {\bibfnamefont {I.}~\bibnamefont
  {Kulik}}\ and\ \bibinfo {author} {\bibfnamefont {A.~N.}\ \bibnamefont
  {Omel'yanchuk}},\ }\href@noop {} {\bibfield  {journal} {\bibinfo  {journal}
  {Soviet Journal of Low Temperature Physics}\ }\textbf {\bibinfo {volume}
  {4}},\ \bibinfo {pages} {142} (\bibinfo {year} {1978})}\BibitemShut {NoStop}%
\bibitem [{\citenamefont {Ambegaokar}\ and\ \citenamefont
  {Baratoff}(1963)}]{amb}%
  \BibitemOpen
  \bibfield  {author} {\bibinfo {author} {\bibfnamefont {V.}~\bibnamefont
  {Ambegaokar}}\ and\ \bibinfo {author} {\bibfnamefont {A.}~\bibnamefont
  {Baratoff}},\ }\href {\doibase 10.1103/PhysRevLett.10.486} {\bibfield
  {journal} {\bibinfo  {journal} {Phys. Rev. Lett.}\ }\textbf {\bibinfo
  {volume} {10}},\ \bibinfo {pages} {486} (\bibinfo {year} {1963})}\BibitemShut
  {NoStop}%
\bibitem [{\citenamefont {Furusaki}(1991)}]{Furusaki1991}%
  \BibitemOpen
  \bibfield  {author} {\bibinfo {author} {\bibfnamefont {M.}~\bibnamefont
  {Furusaki}, \bibfnamefont {A.~andTsukada}},\ }\href@noop {} {\bibfield
  {journal} {\bibinfo  {journal} {Solid State Communications,}\ }\textbf
  {\bibinfo {volume} {78}},\ \bibinfo {pages} {299} (\bibinfo {year}
  {1991})}\BibitemShut {NoStop}%
\bibitem [{\citenamefont {Bagwell}(1992)}]{Bagwell1992}%
  \BibitemOpen
  \bibfield  {author} {\bibinfo {author} {\bibfnamefont {P.~F.}\ \bibnamefont
  {Bagwell}},\ }\href {\doibase 10.1103/PhysRevB.46.12573} {\bibfield
  {journal} {\bibinfo  {journal} {Phys. Rev. B}\ }\textbf {\bibinfo {volume}
  {46}},\ \bibinfo {pages} {12573} (\bibinfo {year} {1992})}\BibitemShut
  {NoStop}%
\bibitem [{\citenamefont {Tanaka}\ and\ \citenamefont
  {Kashiwaya}(1996)}]{tanaka96}%
  \BibitemOpen
  \bibfield  {author} {\bibinfo {author} {\bibfnamefont {Y.}~\bibnamefont
  {Tanaka}}\ and\ \bibinfo {author} {\bibfnamefont {S.}~\bibnamefont
  {Kashiwaya}},\ }\href {\doibase 10.1103/PhysRevB.53.R11957} {\bibfield
  {journal} {\bibinfo  {journal} {Phys. Rev. B}\ }\textbf {\bibinfo {volume}
  {53}},\ \bibinfo {pages} {R11957} (\bibinfo {year} {1996})}\BibitemShut
  {NoStop}%
\bibitem [{\citenamefont {Chang}\ and\ \citenamefont {Bagwell}(1994)}]{bag1}%
  \BibitemOpen
  \bibfield  {author} {\bibinfo {author} {\bibfnamefont {L.-F.}\ \bibnamefont
  {Chang}}\ and\ \bibinfo {author} {\bibfnamefont {P.~F.}\ \bibnamefont
  {Bagwell}},\ }\href {\doibase 10.1103/PhysRevB.49.15853} {\bibfield
  {journal} {\bibinfo  {journal} {Phys. Rev. B}\ }\textbf {\bibinfo {volume}
  {49}},\ \bibinfo {pages} {15853} (\bibinfo {year} {1994})}\BibitemShut
  {NoStop}%
\bibitem [{\citenamefont {Raghu}\ \emph {et~al.}(2008)\citenamefont {Raghu},
  \citenamefont {Qi}, \citenamefont {Liu}, \citenamefont {Scalapino},\ and\
  \citenamefont {Zhang}}]{rag}%
  \BibitemOpen
  \bibfield  {author} {\bibinfo {author} {\bibfnamefont {S.}~\bibnamefont
  {Raghu}}, \bibinfo {author} {\bibfnamefont {X.-L.}\ \bibnamefont {Qi}},
  \bibinfo {author} {\bibfnamefont {C.-X.}\ \bibnamefont {Liu}}, \bibinfo
  {author} {\bibfnamefont {D.~J.}\ \bibnamefont {Scalapino}}, \ and\ \bibinfo
  {author} {\bibfnamefont {S.-C.}\ \bibnamefont {Zhang}},\ }\href {\doibase
  10.1103/PhysRevB.77.220503} {\bibfield  {journal} {\bibinfo  {journal} {Phys.
  Rev. B}\ }\textbf {\bibinfo {volume} {77}},\ \bibinfo {pages} {220503}
  (\bibinfo {year} {2008})}\BibitemShut {NoStop}%
\bibitem [{\citenamefont {Mazin}\ \emph {et~al.}(2008)\citenamefont {Mazin},
  \citenamefont {Singh}, \citenamefont {Johannes},\ and\ \citenamefont
  {Du}}]{maz1}%
  \BibitemOpen
  \bibfield  {author} {\bibinfo {author} {\bibfnamefont {I.~I.}\ \bibnamefont
  {Mazin}}, \bibinfo {author} {\bibfnamefont {D.~J.}\ \bibnamefont {Singh}},
  \bibinfo {author} {\bibfnamefont {M.~D.}\ \bibnamefont {Johannes}}, \ and\
  \bibinfo {author} {\bibfnamefont {M.~H.}\ \bibnamefont {Du}},\ }\href
  {\doibase 10.1103/PhysRevLett.101.057003} {\bibfield  {journal} {\bibinfo
  {journal} {Phys. Rev. Lett.}\ }\textbf {\bibinfo {volume} {101}},\ \bibinfo
  {pages} {057003} (\bibinfo {year} {2008})}\BibitemShut {NoStop}%
\bibitem [{\citenamefont {Kuroki}\ \emph {et~al.}(2008)\citenamefont {Kuroki},
  \citenamefont {Onari}, \citenamefont {Arita}, \citenamefont {Usui},
  \citenamefont {Tanaka}, \citenamefont {Kontani},\ and\ \citenamefont
  {Aoki}}]{Kuroki}%
  \BibitemOpen
  \bibfield  {author} {\bibinfo {author} {\bibfnamefont {K.}~\bibnamefont
  {Kuroki}}, \bibinfo {author} {\bibfnamefont {S.}~\bibnamefont {Onari}},
  \bibinfo {author} {\bibfnamefont {R.}~\bibnamefont {Arita}}, \bibinfo
  {author} {\bibfnamefont {H.}~\bibnamefont {Usui}}, \bibinfo {author}
  {\bibfnamefont {Y.}~\bibnamefont {Tanaka}}, \bibinfo {author} {\bibfnamefont
  {H.}~\bibnamefont {Kontani}}, \ and\ \bibinfo {author} {\bibfnamefont
  {H.}~\bibnamefont {Aoki}},\ }\href {\doibase 10.1103/PhysRevLett.101.087004}
  {\bibfield  {journal} {\bibinfo  {journal} {Phys. Rev. Lett.}\ }\textbf
  {\bibinfo {volume} {101}},\ \bibinfo {pages} {087004} (\bibinfo {year}
  {2008})}\BibitemShut {NoStop}%
\bibitem [{\citenamefont {Kontani}\ and\ \citenamefont
  {Onari}(2010)}]{Kontani}%
  \BibitemOpen
  \bibfield  {author} {\bibinfo {author} {\bibfnamefont {H.}~\bibnamefont
  {Kontani}}\ and\ \bibinfo {author} {\bibfnamefont {S.}~\bibnamefont
  {Onari}},\ }\href {\doibase 10.1103/PhysRevLett.104.157001} {\bibfield
  {journal} {\bibinfo  {journal} {Phys. Rev. Lett.}\ }\textbf {\bibinfo
  {volume} {104}},\ \bibinfo {pages} {157001} (\bibinfo {year}
  {2010})}\BibitemShut {NoStop}%
\bibitem [{\citenamefont {Zhu}\ and\ \citenamefont {Kroemer}(1983)}]{Zhu1983}%
  \BibitemOpen
  \bibfield  {author} {\bibinfo {author} {\bibfnamefont {Q.-G.}\ \bibnamefont
  {Zhu}}\ and\ \bibinfo {author} {\bibfnamefont {H.}~\bibnamefont {Kroemer}},\
  }\href {\doibase 10.1103/PhysRevB.27.3519} {\bibfield  {journal} {\bibinfo
  {journal} {Phys. Rev. B}\ }\textbf {\bibinfo {volume} {27}},\ \bibinfo
  {pages} {3519} (\bibinfo {year} {1983})}\BibitemShut {NoStop}%
\bibitem [{\citenamefont {Moreo}\ \emph {et~al.}(2009)\citenamefont {Moreo},
  \citenamefont {Daghofer}, \citenamefont {Riera},\ and\ \citenamefont
  {Dagotto}}]{mor}%
  \BibitemOpen
  \bibfield  {author} {\bibinfo {author} {\bibfnamefont {A.}~\bibnamefont
  {Moreo}}, \bibinfo {author} {\bibfnamefont {M.}~\bibnamefont {Daghofer}},
  \bibinfo {author} {\bibfnamefont {J.~A.}\ \bibnamefont {Riera}}, \ and\
  \bibinfo {author} {\bibfnamefont {E.}~\bibnamefont {Dagotto}},\ }\href
  {\doibase 10.1103/PhysRevB.79.134502} {\bibfield  {journal} {\bibinfo
  {journal} {Phys. Rev. B}\ }\textbf {\bibinfo {volume} {79}},\ \bibinfo
  {pages} {134502} (\bibinfo {year} {2009})}\BibitemShut {NoStop}%
\bibitem [{\citenamefont {Siedel}(2011)}]{Siedel2011}%
  \BibitemOpen
  \bibfield  {author} {\bibinfo {author} {\bibfnamefont {P.}~\bibnamefont
  {Siedel}},\ }\href@noop {} {\bibfield  {journal} {\bibinfo  {journal}
  {Superconductor Science and Technology}\ }\textbf {\bibinfo {volume} {24}},\
  \bibinfo {pages} {043001} (\bibinfo {year} {2011})}\BibitemShut {NoStop}%
\bibitem [{\citenamefont {Liu}\ \emph {et~al.}(2014)\citenamefont {Liu},
  \citenamefont {Tanatar}, \citenamefont {Straszheim}, \citenamefont {Jensen},
  \citenamefont {Dennis}, \citenamefont {McCallum}, \citenamefont {Kogan},
  \citenamefont {Prozorov},\ and\ \citenamefont {Lograsso}}]{Liu}%
  \BibitemOpen
  \bibfield  {author} {\bibinfo {author} {\bibfnamefont {Y.}~\bibnamefont
  {Liu}}, \bibinfo {author} {\bibfnamefont {M.~A.}\ \bibnamefont {Tanatar}},
  \bibinfo {author} {\bibfnamefont {W.~E.}\ \bibnamefont {Straszheim}},
  \bibinfo {author} {\bibfnamefont {B.}~\bibnamefont {Jensen}}, \bibinfo
  {author} {\bibfnamefont {K.~W.}\ \bibnamefont {Dennis}}, \bibinfo {author}
  {\bibfnamefont {R.~W.}\ \bibnamefont {McCallum}}, \bibinfo {author}
  {\bibfnamefont {V.~G.}\ \bibnamefont {Kogan}}, \bibinfo {author}
  {\bibfnamefont {R.}~\bibnamefont {Prozorov}}, \ and\ \bibinfo {author}
  {\bibfnamefont {T.~A.}\ \bibnamefont {Lograsso}},\ }\href {\doibase
  10.1103/PhysRevB.89.134504} {\bibfield  {journal} {\bibinfo  {journal} {Phys.
  Rev. B}\ }\textbf {\bibinfo {volume} {89}},\ \bibinfo {pages} {134504}
  (\bibinfo {year} {2014})}\BibitemShut {NoStop}%
\bibitem [{\citenamefont {Avci}\ \emph {et~al.}(2012)\citenamefont {Avci},
  \citenamefont {Chmaissem}, \citenamefont {Chung}, \citenamefont {Rosenkranz},
  \citenamefont {Goremychkin}, \citenamefont {Castellan}, \citenamefont
  {Todorov}, \citenamefont {Schlueter}, \citenamefont {Claus}, \citenamefont
  {Daoud-Aladine}, \citenamefont {Khalyavin}, \citenamefont {Kanatzidis},\ and\
  \citenamefont {Osborn}}]{Avci}%
  \BibitemOpen
  \bibfield  {author} {\bibinfo {author} {\bibfnamefont {S.}~\bibnamefont
  {Avci}}, \bibinfo {author} {\bibfnamefont {O.}~\bibnamefont {Chmaissem}},
  \bibinfo {author} {\bibfnamefont {D.~Y.}\ \bibnamefont {Chung}}, \bibinfo
  {author} {\bibfnamefont {S.}~\bibnamefont {Rosenkranz}}, \bibinfo {author}
  {\bibfnamefont {E.~A.}\ \bibnamefont {Goremychkin}}, \bibinfo {author}
  {\bibfnamefont {J.~P.}\ \bibnamefont {Castellan}}, \bibinfo {author}
  {\bibfnamefont {I.~S.}\ \bibnamefont {Todorov}}, \bibinfo {author}
  {\bibfnamefont {J.~A.}\ \bibnamefont {Schlueter}}, \bibinfo {author}
  {\bibfnamefont {H.}~\bibnamefont {Claus}}, \bibinfo {author} {\bibfnamefont
  {A.}~\bibnamefont {Daoud-Aladine}}, \bibinfo {author} {\bibfnamefont {D.~D.}\
  \bibnamefont {Khalyavin}}, \bibinfo {author} {\bibfnamefont {M.~G.}\
  \bibnamefont {Kanatzidis}}, \ and\ \bibinfo {author} {\bibfnamefont
  {R.}~\bibnamefont {Osborn}},\ }\href {\doibase 10.1103/PhysRevB.85.184507}
  {\bibfield  {journal} {\bibinfo  {journal} {Phys. Rev. B}\ }\textbf {\bibinfo
  {volume} {85}},\ \bibinfo {pages} {184507} (\bibinfo {year}
  {2012})}\BibitemShut {NoStop}%
\bibitem [{\citenamefont {{Barone, A. and Patern\`{o}, G.}}(1982)}]{barone}%
  \BibitemOpen
  \bibfield  {author} {\bibinfo {author} {\bibnamefont {{Barone, A. and
  Patern\`{o}, G.}}},\ }\href@noop {} {\emph {\bibinfo {title} {Physics and
  Applications of the Josephson Effect}}}\ (\bibinfo  {publisher} {John Wiley
  \& Sons},\ \bibinfo {year} {1982})\BibitemShut {NoStop}%
\bibitem [{\citenamefont {Sellier}\ \emph {et~al.}(2004)\citenamefont
  {Sellier}, \citenamefont {Baraduc}, \citenamefont {Lefloch},\ and\
  \citenamefont {Calemczuk}}]{Sellier}%
  \BibitemOpen
  \bibfield  {author} {\bibinfo {author} {\bibfnamefont {H.}~\bibnamefont
  {Sellier}}, \bibinfo {author} {\bibfnamefont {C.}~\bibnamefont {Baraduc}},
  \bibinfo {author} {\bibfnamefont {F.}~\bibnamefont {Lefloch}}, \ and\
  \bibinfo {author} {\bibfnamefont {R.}~\bibnamefont {Calemczuk}},\ }\href@noop
  {} {\bibfield  {journal} {\bibinfo  {journal} {Phys. Rev. Lett.}\ }\textbf
  {\bibinfo {volume} {92}},\ \bibinfo {pages} {257005} (\bibinfo {year}
  {2004})}\BibitemShut {NoStop}%
\bibitem [{\citenamefont {Kleiner}\ \emph {et~al.}(1996)\citenamefont
  {Kleiner}, \citenamefont {Katz}, \citenamefont {Sun}, \citenamefont {Summer},
  \citenamefont {Gajewski}, \citenamefont {Han}, \citenamefont {Woods},
  \citenamefont {Dantsker}, \citenamefont {Chen}, \citenamefont {Char},
  \citenamefont {Maple}, \citenamefont {Dynes},\ and\ \citenamefont
  {Clarke}}]{Kleiner}%
  \BibitemOpen
  \bibfield  {author} {\bibinfo {author} {\bibfnamefont {R.}~\bibnamefont
  {Kleiner}}, \bibinfo {author} {\bibfnamefont {A.~S.}\ \bibnamefont {Katz}},
  \bibinfo {author} {\bibfnamefont {A.~G.}\ \bibnamefont {Sun}}, \bibinfo
  {author} {\bibfnamefont {R.}~\bibnamefont {Summer}}, \bibinfo {author}
  {\bibfnamefont {D.~A.}\ \bibnamefont {Gajewski}}, \bibinfo {author}
  {\bibfnamefont {S.~H.}\ \bibnamefont {Han}}, \bibinfo {author} {\bibfnamefont
  {S.~I.}\ \bibnamefont {Woods}}, \bibinfo {author} {\bibfnamefont
  {E.}~\bibnamefont {Dantsker}}, \bibinfo {author} {\bibfnamefont
  {B.}~\bibnamefont {Chen}}, \bibinfo {author} {\bibfnamefont {K.}~\bibnamefont
  {Char}}, \bibinfo {author} {\bibfnamefont {M.~B.}\ \bibnamefont {Maple}},
  \bibinfo {author} {\bibfnamefont {R.~C.}\ \bibnamefont {Dynes}}, \ and\
  \bibinfo {author} {\bibfnamefont {J.}~\bibnamefont {Clarke}},\ }\href@noop {}
  {\bibfield  {journal} {\bibinfo  {journal} {Phys. Rev. Lett.}\ }\textbf
  {\bibinfo {volume} {76}},\ \bibinfo {pages} {2161} (\bibinfo {year}
  {1996})}\BibitemShut {NoStop}%
\bibitem [{\citenamefont {Belenov}\ \emph {et~al.}(1979)\citenamefont
  {Belenov}, \citenamefont {Vedeneev}, \citenamefont {Motulevich},
  \citenamefont {Stepanov},\ and\ \citenamefont {Uskov}}]{belenov}%
  \BibitemOpen
  \bibfield  {author} {\bibinfo {author} {\bibfnamefont {E.~M.}\ \bibnamefont
  {Belenov}}, \bibinfo {author} {\bibfnamefont {S.~I.}\ \bibnamefont
  {Vedeneev}}, \bibinfo {author} {\bibfnamefont {G.~P.}\ \bibnamefont
  {Motulevich}}, \bibinfo {author} {\bibfnamefont {V.~A.}\ \bibnamefont
  {Stepanov}}, \ and\ \bibinfo {author} {\bibfnamefont {A.~V.}\ \bibnamefont
  {Uskov}},\ }\href@noop {} {\bibfield  {journal} {\bibinfo  {journal} {JETP
  Letters}\ }\textbf {\bibinfo {volume} {49}},\ \bibinfo {pages} {399}
  (\bibinfo {year} {1979})}\BibitemShut {NoStop}%
\bibitem [{\citenamefont {Seidel}\ \emph {et~al.}(1991)\citenamefont {Seidel},
  \citenamefont {Siegel},\ and\ \citenamefont {Heinz}}]{Seidel}%
  \BibitemOpen
  \bibfield  {author} {\bibinfo {author} {\bibfnamefont {P.}~\bibnamefont
  {Seidel}}, \bibinfo {author} {\bibfnamefont {M.}~\bibnamefont {Siegel}}, \
  and\ \bibinfo {author} {\bibfnamefont {E.}~\bibnamefont {Heinz}},\
  }\href@noop {} {\bibfield  {journal} {\bibinfo  {journal} {Physica C}\
  }\textbf {\bibinfo {volume} {180}},\ \bibinfo {pages} {284} (\bibinfo {year}
  {1991})}\BibitemShut {NoStop}%
\bibitem [{\citenamefont {Cuevas}\ \emph {et~al.}(2002)\citenamefont {Cuevas},
  \citenamefont {Heurich}, \citenamefont {Mart\'{\i}n-Rodero}, \citenamefont
  {Levy~Yeyati},\ and\ \citenamefont {Sch\"{o}n}}]{cuevas}%
  \BibitemOpen
  \bibfield  {author} {\bibinfo {author} {\bibfnamefont {J.~C.}\ \bibnamefont
  {Cuevas}}, \bibinfo {author} {\bibfnamefont {J.}~\bibnamefont {Heurich}},
  \bibinfo {author} {\bibfnamefont {A.}~\bibnamefont {Mart\'{\i}n-Rodero}},
  \bibinfo {author} {\bibfnamefont {A.}~\bibnamefont {Levy~Yeyati}}, \ and\
  \bibinfo {author} {\bibfnamefont {G.}~\bibnamefont {Sch\"{o}n}},\ }\href@noop
  {} {\bibfield  {journal} {\bibinfo  {journal} {Phys. Rev. Lett.}\ }\textbf
  {\bibinfo {volume} {88}},\ \bibinfo {pages} {157001} (\bibinfo {year}
  {2002})}\BibitemShut {NoStop}%
\bibitem [{\citenamefont {Yada}\ \emph {et~al.}(2014)\citenamefont {Yada},
  \citenamefont {Golubov}, \citenamefont {Tanaka},\ and\ \citenamefont
  {Kashiwaya}}]{Yada2014}%
  \BibitemOpen
  \bibfield  {author} {\bibinfo {author} {\bibfnamefont {K.}~\bibnamefont
  {Yada}}, \bibinfo {author} {\bibfnamefont {A.~A.}\ \bibnamefont {Golubov}},
  \bibinfo {author} {\bibfnamefont {Y.}~\bibnamefont {Tanaka}}, \ and\ \bibinfo
  {author} {\bibfnamefont {S.}~\bibnamefont {Kashiwaya}},\ }\href@noop {}
  {\bibfield  {journal} {\bibinfo  {journal} {Journal of the Physical Society
  of Japan}\ }\textbf {\bibinfo {volume} {83}},\ \bibinfo {pages} {074706}
  (\bibinfo {year} {2014})}\BibitemShut {NoStop}%
\bibitem [{\citenamefont {Fu}\ and\ \citenamefont {Kane}(2008)}]{fu08}%
  \BibitemOpen
  \bibfield  {author} {\bibinfo {author} {\bibfnamefont {L.}~\bibnamefont
  {Fu}}\ and\ \bibinfo {author} {\bibfnamefont {C.~L.}\ \bibnamefont {Kane}},\
  }\href@noop {} {\bibfield  {journal} {\bibinfo  {journal} {Phys. Rev. Lett.}\
  }\textbf {\bibinfo {volume} {100}},\ \bibinfo {pages} {096407} (\bibinfo
  {year} {2008})}\BibitemShut {NoStop}%
\bibitem [{\citenamefont {Tanaka}\ \emph {et~al.}(2009)\citenamefont {Tanaka},
  \citenamefont {Yokoyama},\ and\ \citenamefont {Nagaosa}}]{tanaka09}%
  \BibitemOpen
  \bibfield  {author} {\bibinfo {author} {\bibfnamefont {Y.}~\bibnamefont
  {Tanaka}}, \bibinfo {author} {\bibfnamefont {T.}~\bibnamefont {Yokoyama}}, \
  and\ \bibinfo {author} {\bibfnamefont {N.}~\bibnamefont {Nagaosa}},\ }\href
  {\doibase 10.1103/PhysRevLett.103.107002} {\bibfield  {journal} {\bibinfo
  {journal} {Phys. Rev. Lett.}\ }\textbf {\bibinfo {volume} {103}},\ \bibinfo
  {pages} {107002} (\bibinfo {year} {2009})}\BibitemShut {NoStop}%
\bibitem [{\citenamefont {Linder}\ \emph {et~al.}(2010)\citenamefont {Linder},
  \citenamefont {Tanaka}, \citenamefont {Yokoyama}, \citenamefont {Sudbo},\
  and\ \citenamefont {Nagaosa}}]{Linder10}%
  \BibitemOpen
  \bibfield  {author} {\bibinfo {author} {\bibfnamefont {J.}~\bibnamefont
  {Linder}}, \bibinfo {author} {\bibfnamefont {Y.}~\bibnamefont {Tanaka}},
  \bibinfo {author} {\bibfnamefont {T.}~\bibnamefont {Yokoyama}}, \bibinfo
  {author} {\bibfnamefont {A.}~\bibnamefont {Sudbo}}, \ and\ \bibinfo {author}
  {\bibfnamefont {N.}~\bibnamefont {Nagaosa}},\ }\href@noop {} {\bibfield
  {journal} {\bibinfo  {journal} {Phys. Rev. Lett.}\ }\textbf {\bibinfo
  {volume} {104}},\ \bibinfo {pages} {067001} (\bibinfo {year}
  {2010})}\BibitemShut {NoStop}%
\bibitem [{\citenamefont {Yamakage}\ \emph {et~al.}(2012)\citenamefont
  {Yamakage}, \citenamefont {Yada}, \citenamefont {Sato},\ and\ \citenamefont
  {Tanaka}}]{yamakage12}%
  \BibitemOpen
  \bibfield  {author} {\bibinfo {author} {\bibfnamefont {A.}~\bibnamefont
  {Yamakage}}, \bibinfo {author} {\bibfnamefont {K.}~\bibnamefont {Yada}},
  \bibinfo {author} {\bibfnamefont {M.}~\bibnamefont {Sato}}, \ and\ \bibinfo
  {author} {\bibfnamefont {Y.}~\bibnamefont {Tanaka}},\ }\href {\doibase
  10.1103/PhysRevB.85.180509} {\bibfield  {journal} {\bibinfo  {journal} {Phys.
  Rev. B}\ }\textbf {\bibinfo {volume} {85}},\ \bibinfo {pages} {180509}
  (\bibinfo {year} {2012})}\BibitemShut {NoStop}%
\bibitem [{\citenamefont {Tanaka}\ and\ \citenamefont
  {Golubov}(2007)}]{Tanaka2007}%
  \BibitemOpen
  \bibfield  {author} {\bibinfo {author} {\bibfnamefont {Y.}~\bibnamefont
  {Tanaka}}\ and\ \bibinfo {author} {\bibfnamefont {A.~A.}\ \bibnamefont
  {Golubov}},\ }\href@noop {} {\bibfield  {journal} {\bibinfo  {journal} {Phys.
  Rev. Lett.}\ }\textbf {\bibinfo {volume} {98}},\ \bibinfo {pages} {037003}
  (\bibinfo {year} {2007})}\BibitemShut {NoStop}%
\bibitem [{\citenamefont {Tanaka}\ \emph {et~al.}(2012)\citenamefont {Tanaka},
  \citenamefont {Sato},\ and\ \citenamefont {Nagaosa}}]{tanaka12}%
  \BibitemOpen
  \bibfield  {author} {\bibinfo {author} {\bibfnamefont {Y.}~\bibnamefont
  {Tanaka}}, \bibinfo {author} {\bibfnamefont {M.}~\bibnamefont {Sato}}, \ and\
  \bibinfo {author} {\bibfnamefont {N.}~\bibnamefont {Nagaosa}},\ }\href
  {\doibase 10.1143/JPSJ.81.011013} {\bibfield  {journal} {\bibinfo  {journal}
  {J. Phys. Soc. Jpn.}\ }\textbf {\bibinfo {volume} {81}},\ \bibinfo {pages}
  {011013} (\bibinfo {year} {2012})}\BibitemShut {NoStop}%
\bibitem [{\citenamefont {Black-Schaffer}\ and\ \citenamefont
  {Balatsky}(2013{\natexlab{a}})}]{Balatsky1}%
  \BibitemOpen
  \bibfield  {author} {\bibinfo {author} {\bibfnamefont {A.~M.}\ \bibnamefont
  {Black-Schaffer}}\ and\ \bibinfo {author} {\bibfnamefont {A.~V.}\
  \bibnamefont {Balatsky}},\ }\href@noop {} {\bibfield  {journal} {\bibinfo
  {journal} {Phys. Rev. B}\ }\textbf {\bibinfo {volume} {88}},\ \bibinfo
  {pages} {104514} (\bibinfo {year} {2013}{\natexlab{a}})}\BibitemShut
  {NoStop}%
\bibitem [{\citenamefont {Black-Schaffer}\ and\ \citenamefont
  {Balatsky}(2013{\natexlab{b}})}]{Balatsky2}%
  \BibitemOpen
  \bibfield  {author} {\bibinfo {author} {\bibfnamefont {A.~M.}\ \bibnamefont
  {Black-Schaffer}}\ and\ \bibinfo {author} {\bibfnamefont {A.~V.}\
  \bibnamefont {Balatsky}},\ }\href@noop {} {\bibfield  {journal} {\bibinfo
  {journal} {Phys. Rev. B}\ }\textbf {\bibinfo {volume} {87}},\ \bibinfo
  {pages} {220506} (\bibinfo {year} {2013}{\natexlab{b}})}\BibitemShut
  {NoStop}%
\bibitem [{\citenamefont {Hao}\ \emph {et~al.}(2011)\citenamefont {Hao},
  \citenamefont {Thalmeier},\ and\ \citenamefont {Lee}}]{Hao1}%
  \BibitemOpen
  \bibfield  {author} {\bibinfo {author} {\bibfnamefont {L.}~\bibnamefont
  {Hao}}, \bibinfo {author} {\bibfnamefont {P.}~\bibnamefont {Thalmeier}}, \
  and\ \bibinfo {author} {\bibfnamefont {T.~K.}\ \bibnamefont {Lee}},\
  }\href@noop {} {\bibfield  {journal} {\bibinfo  {journal} {Phys. Rev. B}\
  }\textbf {\bibinfo {volume} {84}},\ \bibinfo {pages} {235303} (\bibinfo
  {year} {2011})}\BibitemShut {NoStop}%
\bibitem [{\citenamefont {Hao}\ \emph {et~al.}(2014)\citenamefont {Hao},
  \citenamefont {Wang}, \citenamefont {Lee}, \citenamefont {Wang},
  \citenamefont {Tsai},\ and\ \citenamefont {Yang}}]{Hao2}%
  \BibitemOpen
  \bibfield  {author} {\bibinfo {author} {\bibfnamefont {L.}~\bibnamefont
  {Hao}}, \bibinfo {author} {\bibfnamefont {G.-L.}\ \bibnamefont {Wang}},
  \bibinfo {author} {\bibfnamefont {T.-K.}\ \bibnamefont {Lee}}, \bibinfo
  {author} {\bibfnamefont {J.}~\bibnamefont {Wang}}, \bibinfo {author}
  {\bibfnamefont {W.-F.}\ \bibnamefont {Tsai}}, \ and\ \bibinfo {author}
  {\bibfnamefont {Y.-H.}\ \bibnamefont {Yang}},\ }\href@noop {} {\bibfield
  {journal} {\bibinfo  {journal} {Phys. Rev. B}\ }\textbf {\bibinfo {volume}
  {89}},\ \bibinfo {pages} {214505} (\bibinfo {year} {2014})}\BibitemShut
  {NoStop}%
\end{thebibliography}%

\end{document}